\documentclass[aps,pra,reprint,superscriptaddress,showpacs]{revtex4-1}

\usepackage{graphicx}
\usepackage{dcolumn}
\usepackage{bm}
\usepackage{xcolor}
\usepackage{scrextend}		
\usepackage{url}
\usepackage{epstopdf}
\usepackage{hyperref}
\usepackage{subfigure}
\usepackage{amsmath,bm} 
\usepackage{gensymb} 
\usepackage{url}
\usepackage{hyperref}

\begin{document}

\newcommand{\jmb}[1]{\textcolor{blue}{#1}}
\newcommand{\jmr}[1]{\textcolor{red}{#1}}
\newcommand{\mmv}[1]{\textcolor{violet}{#1}}
\newcommand{\hg}{HgCr$_2$Se$_4$}

\title{Probing the magnetic polaron state in the ferromagnetic semiconductor HgCr$_2$Se$_4$ with resistance fluctuation and muon-spin spectroscopy measurements}

\author{Merlin~Mitschek}
\affiliation{Institute of Physics, Goethe-University Frankfurt, 60438 Frankfurt (Main), Germany}
\author{Thomas~J.~Hicken}
\altaffiliation{Current address: Department of Physics, Royal Holloway, University of London, Egham TW20 0EX, United Kingdom}
\affiliation{Department of Physics, Centre for Materials Physics, Durham University, Durham, DH1 3LE, United Kingdom}
\author{Shuai~Yang}
\affiliation{Beijing National Laboratory for Condensed Matter Physics, Institute of Physics, Chinese Academy of Sciences, Beijing 100190, China}
\author{Murray~N.~Wilson}
\affiliation{Department of Physics, Centre for Materials Physics, Durham University, Durham, DH1 3LE, United Kingdom}
\author{Francis~L.~Pratt}
\affiliation{ISIS Pulsed Neutron and Muon Facility, STFC Rutherford Appleton Laboratory, Harwell Oxford, Didcot, OX11 OQX, United Kingdom}
\author{Chennan~Wang}
\affiliation{Laboratory for Muon-Spin Spectroscopy, Paul Scherrer Institut, Forschungsstrasse 111, 5232 Villigen PSI, Switzerland}
\author{Stephen~J.~Blundell}
\affiliation{Department of Physics, Clarendon Laboratory, Oxford University, Parks Road, Oxford OX1 3PU, United Kingdom}
\author{Zhilin~Li}
\affiliation{Beijing National Laboratory for Condensed Matter Physics, Institute of Physics, Chinese Academy of Sciences, Beijing 100190, China}
\author{Yongqing~Li}
\affiliation{Beijing National Laboratory for Condensed Matter Physics, Institute of Physics, Chinese Academy of Sciences, Beijing 100190, China}
\author{Tom~Lancaster}
\affiliation{Department of Physics, Centre for Materials Physics, Durham University, Durham, DH1 3LE, United Kingdom}
\author{Jens~M\"uller}
\email{j.mueller@physik.uni-frankfurt.de}
\affiliation{Institute of Physics, Goethe-University Frankfurt, 60438 Frankfurt (Main), Germany}

\date{\today}

\begin{abstract}
Combined resistance noise and muon-spin relaxation ($\mu$SR) measurements of the ferromagnetic semiconductor HgCr$_2$Se$_4$ suggest a degree of magnetoelectric coupling and provide evidence for the existence of isolated magnetic polarons. These form at elevated temperatures and undergo a percolation transition with a drastic enhancement of the low-frequency 1/$f$-type charge fluctuations at the insulator-to-metal transition at $\sim 95 - 98$\,K in the vicinity of the magnetic ordering temperature $T_C \sim 105 - 107$\,K. Upon approaching the percolation threshold from above, the strikingly unusual dynamics of a distinct two-level fluctuator superimposed on the $1/f$ noise can be described by a slowing down of the dynamics of a nanoscale magnetic cluster, a magnetic polaron, when taking into account an effective radius of the polaron depending on the spin correlation length. Coinciding temperature scales found in $\mu$SR and noise measurements suggest changes in the magnetic dynamics over a wide range of frequencies and are consistent with the existence of large polarized and domain-wall-like regions at low temperatures, that result from the freezing of spin dynamics at the magnetic polaron percolation transition.
\end{abstract}


\maketitle

\section{Introduction}
HgCr$_2$Se$_4$ is a magnetic semiconductor that forms part of the larger group of chromium chalcogenides, which are spinels with formula ACr$_2$X$_4$. The ferromagnetic interactions in HgCr$_2$Se$_4$ are so strong that the antiferromagnetic exchange, often dominant for the spinel structure, is overwhelmed and ferromagnetism is observed \cite{Baltzer1966}.
Recently HgCr$_2$Se$_4$ has attracted attention when doped with $n$-type impurities, both as a half metal and potential Chern semimetal \cite{Guan2015}, exhibiting unconventional low-temperature transport properties \cite{Yang2019}, and due to the colossal magnetoresistance (CMR) effect, which is observed in the vicinity of the magnetic ordering temperature $T_C \simeq 105$\,K \cite{Lin2016}. The magnetic ordering drives an insulator-to-metal transition. Whilst the bulk magnetic properties of the system are well characterized, with magnetization measurements demonstrating it has critical behavior very close to that of an ideal 3D Heisenberg magnet \cite{Lin2016}, the spin fluctuations are not so well studied. Understanding these fluctuations is particularly important, as spin correlations have been proposed to play an important role in the transport properties not only near the critical point but also for a wide range of temperatures $T_C < T < T^\ast$ above the ferromagnetic transition in the paramagnetic phase \cite{Lin2016}. Likewise, charge fluctuations which are often linked to magnetically-driven electronic phase separation and magnetic cluster formation, and which frequently are invoked as ingredients of models explaining the CMR effect, have not been investigated. 
Nanoscale ferromagnetic clusters, magnetic polarons (MP), may form due to the large exchange interaction between charge carriers and localized spins --- and become enhanced by spin correlations, an effect that has been largely neglected in previous studies. Materials prone to the formation of such nanoscale spin-charge composites are systems with low carrier density where a localized carrier (non-ionized donors, in HgCr$_2$Se$_4$ presumably Se vacancies) polarizes the surrounding spins in the lattice of magnetic ions \cite{Molnar2007}. A percolation transition of MPs has been suggested to cause the observed CMR effect in a variety of materials \cite{Majumdar1998,Molnar2007}, as e.g.\ in perovskite manganites \cite{Teresa1997,Faeth1999,Adams2000,Tokura2006}, EuB$_6$ \cite{Nyhus1997,Suellow1998,Suellow2000,Urbano2004,Yu2006,Zhang2009}, and also in the family of HgCr$_2$Se$_4$ \cite{Yang2004,Lin2016,Li2018,Xia2019}. Here, a change of electronic transport regimes and the deviation of the magnetic susceptibility from the Curie-Weiss law at a temperature $T^\ast  \sim 2\,T_C$ is associated with the formation of isolated MPs which percolate upon lowering the temperature \cite{Lin2016}. A continuous network is formed at the percolation threshold which upon cooling the sample is reached in the vicinity of the magnetic transition or may even coincide with $T_C$. The trapped carriers forming the MPs then become unbound in an external or internal magnetic field which results in a drastic increase of their mobility giving rise to the CMR behavior.

In HgCr$_2$Se$_4$, besides the spin fluctuations, the fluctuation dynamics in the charge sector and the mutual, magnetoelectric coupling remains to be explored. Since for these effects vastly different time scales are expected, we performed muon-spin relaxation ($\mu$SR) experiments which can be used to study dynamic magnetism at frequencies within the range 
 $\sim 10^5 - 10^{11}$\,Hz and resistance noise measurements for studying the charge carrier dynamics at low frequencies $\sim 10^{-2} - 10^2$\,Hz.
We find coinciding characteristic temperature scales suggesting that the magnetic dynamics in this system change over a wide range of frequencies. The dynamics and the observed coupling of the magnetism and the electronic degrees of freedom is consistent with the picture of isolated and diluted magnetic polarons forming at temperatures far above and coalescing near $T_C$, and a network of merged polarons below. In particular, we provide evidence for the previously suggested hypothesis \cite{Lin2016} that in HgCr$_2$Se$_4$ the effective size of the magnetic polarons is dependent on the spin correlation length.

\section{Experimental Details}
\paragraph*{Sample growth}
HgCr$_2$Se$_4$ crystals were grown by chemical vapor transport method using CrCl$_3$ as the main transport agent \cite{Arai1973,Wang2013}. In a typical run, HgCr$_2$Se$_4$ powder (which was synthesized with Hg drops, Cr powder and Se shots) was used as the precursor. The HgCr$_2$Se$_4$ powder and CrCl$_3$ flakes were loaded into a silica ampule under argon, then evacuated to a pressure below 1\,Pa and sealed quickly by flame. The ampule was subsequently exposed to a temperature gradient from 680 to 660$^\circ$C for 2 weeks and naturally cooled down to room temperature. Crystals with sizes 1 -- 3\,mm with regular shapes and shiny facets were collected at the cooler end of the silica tube. The carrier density in the HgCr$_2$Se$_4$ crystals can be adjusted by the growth temperature, the amount of transport agent, and the stoichiometry of source materials. Note that the growth procedure is the same as in \cite{Yang2019} but different from the one in \cite{Lin2016}.\\ 


\paragraph*{$\mu$SR spectroscopy}
In a $\mu$SR experiment~\cite{blundell1999spin} spin-polarized muons are implanted in a sample where they interact with the local magnetic field at the muon site.
After, on average, 2.2~$\mu$s, the muon decays into a positron and two neutrinos. By detecting the positrons, which are preferentially emitted in the direction of the muon spin at the time of decay, we can track the polarisation of the muon-spin ensemble.
In a zero-field (ZF) $\mu$SR experiment, as is primarily considered here, the local magnetic field at the muon sites arises due to the  configuration of the spins in the system.
When the muon-spin has a component perpendicular to the local field $B$, precession occurs with angular frequency $\omega~=~\gamma_\mu B$, where $\gamma_\mu~=~2\pi\times135.5$~MHz~T$^{-1}$ is the gyromagnetic ratio of the muon.
Conversely, when the muon-spin aligns with the local field, only dynamic fluctuations can depolarize the muon-spin ensemble.
The quantity of interest in the experiment is the asymmetry $A\left(t\right)$, calculated from the number of counts arriving in the detectors forwards and backwards of the initial muon-spin polarisation direction, $n_{\rm F,B}$, via $A\left(t\right)=\left(n_{\rm F}-n_{\rm B}\right)/\left(n_{\rm F}+n_{\rm B}\right)$.
The asymmetry is directly proportional to the muon-spin polarisation.
Information on both the static and dynamic magnetism at the muon site can be determined through the functional form of $A\left(t\right)$.

Muon-spin spectroscopy measurements of polycrystalline HgCr$_2$Se$_4$ were carried out at the STFC-ISIS facility, UK, using the EMU instrument, and at the Swiss Muon Source (S$\mu$S), Paul Scherrer Institut, Switzerland, using the GPS instrument.
Polycrystalline samples were packed in Ag-foil packets and loaded in a $^4$He cryostat.
On the EMU instrument the packet was mounted on a silver backing plate, whereas on the GPS instrument the packet was attached to a fork and suspended in the beam, minimising the background contribution to $A\left(t\right)$.
Most measurements were performed in ZF, with additional measurements at ISIS performed in a 2~mT field  applied perpendicular to the initial muon-spin polarisation (known as weak transverse field or wTF measurements).
Data analysis was carried out using the WiMDA program~\cite{pratt2000wimda} and made use of the MINUIT algorithm~\cite{james1975minuit} via the iminuit~\cite{iminuit} Python interface for global refinement of parameters.\\

\begin{figure}[h!]
\centering
\includegraphics[width=\columnwidth]{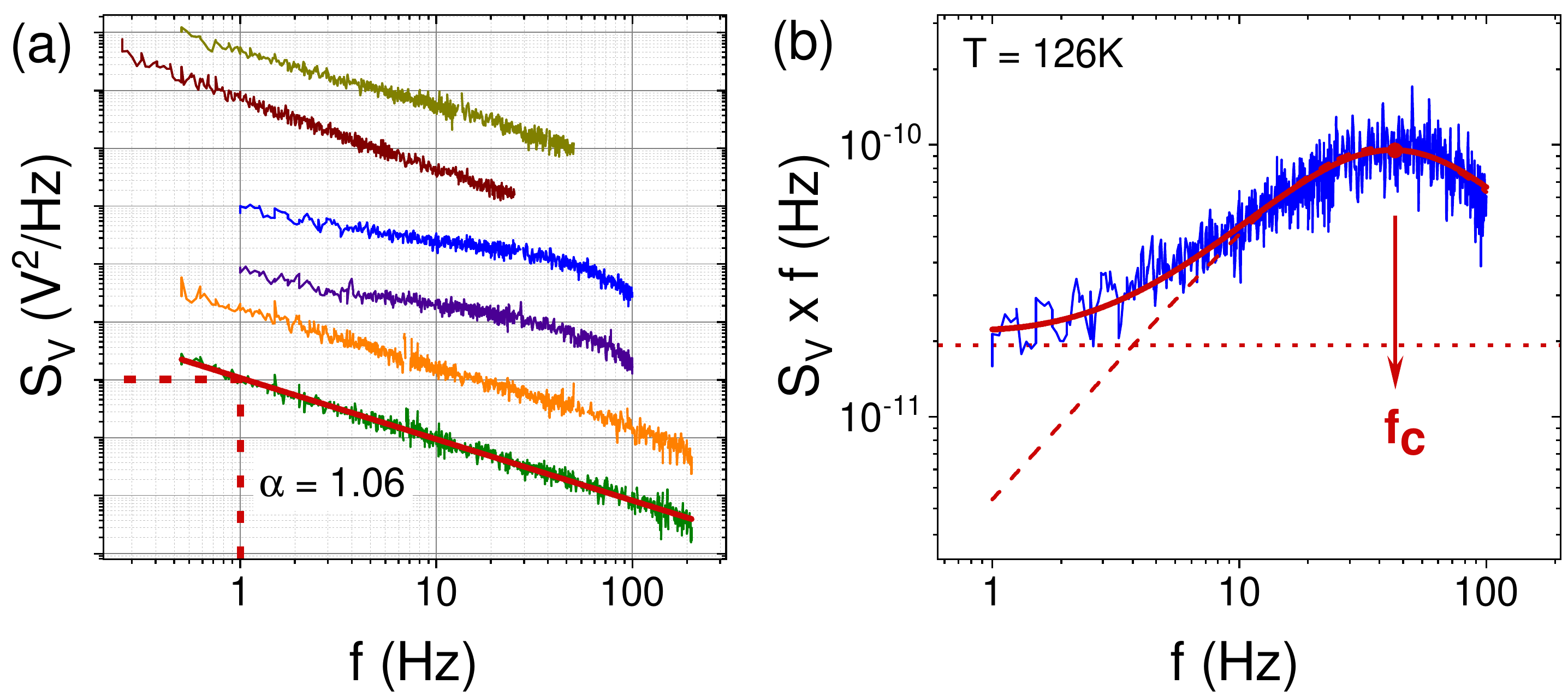}
\caption{Typical voltage noise spectra $S_V$ vs.\ $f$ for HgCr$_2$Se$_4$ (at various temperatures) either of $1/f^\alpha$-type or superimposed with a Lorentzian contribution. (a) For the former type, from a linear fit in the $\log$-$\log$ representation (red line) the magnitude $S_V(f = 1\,{\rm Hz})$ (dashed red line) and the frequency exponent $\alpha$ (slope) is determined. (b) A characteristic spectrum of the latter type shown as $S_V \times f$ vs.\ $f$, where the $1/f^\alpha$ contribution (dotted line) is nearly constant (in this case $\alpha = 1$). The spectrum is fitted to Eq.\,(\ref{noise_spectra}) (continuous red line) with the Lorentzian (dashed line) exhibiting a peak at the corner frequency $f_c$.} 
\label{spectra}
\end{figure}

\paragraph*{Fluctuation (noise) spectroscopy} 
For electrical transport measurements Ti/Au electrical contacts were deposited onto the single crystal HgCr$_2$Se$_4$ samples under a hand-painted photoresist mask in a high vacuum electron beam evaporator. The samples were then annealed in a H/He forming gas in order to ensure good-quality (ohmic) contacts. Gold wires have been attached by silver epoxy, see Supplemental Material of Ref.\,\cite{Yang2019}. 
Single crystals of HgCr$_2$Se$_4$ were measured in a continuous helium-flow cryostat with variable temperature insert equipped with a superconducting magnet, using a four-terminal AC technique and a lock-in amplifier. 
For the resistance noise measurements, after pre-amplification of the voltage drop across the sample, the signal is demodulated by the lock-in amplifier and the voltage noise power spectral density (PSD), $S_V(f,T)$, of the fluctuations is processed and recorded by a spectrum analyzer \cite{Mueller2011,Mueller2018}. As required, the voltage noise PSD scales as $S_V \propto I^2$ and care has been taken to rule out spurious noise sources contributing to the measured signal, e.g.\ from the preamplifier or the electrical contacts. The resistance noise PSD is given by $S_R = S_V/I^2$ and therefore for the normalized quantities it is $S_V/V^2 = S_R/R^2$. Figure \ref{spectra} shows typical spectra of either pure $1/f^\alpha$-type or superimposed with a Lorentzian contribution, the latter being a signature of a two-level switching process. 

\section{Experimental Results} 
\subsection{$\mu$SR spectroscopy}
\begin{figure}[b]
	\centering
	\includegraphics[width=\linewidth]{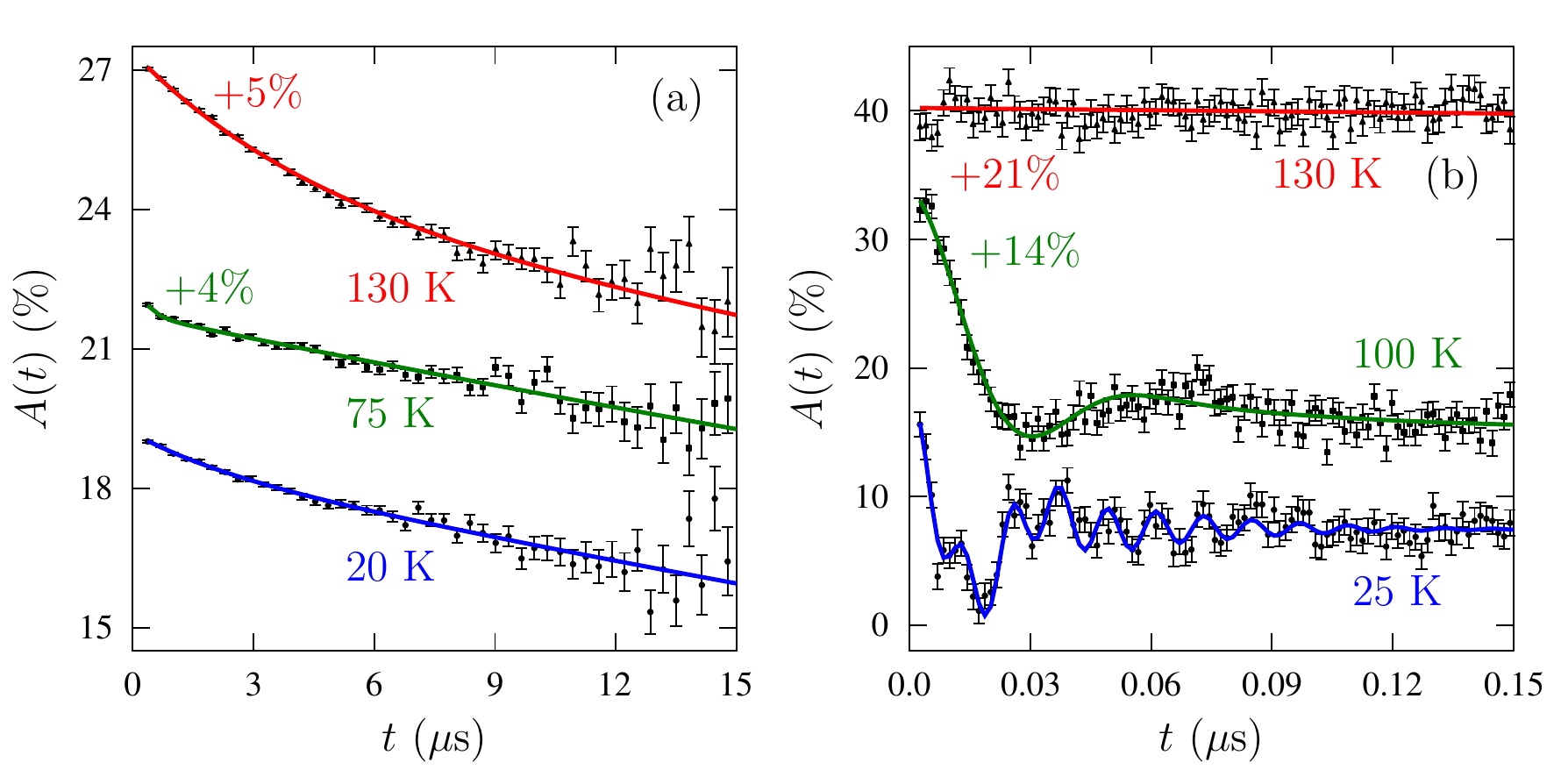}
	\caption{Representative ZF $\mu$SR measurements of HgCr$_2$Se$_4$. Data in (a) was measured at ISIS; (b) shows data measured at S$\mu$S. Fits to the data, described in the text, are shown. Some data are plotted with a labelled vertical offset for clarity.}
	\label{fig:data}
\end{figure}
\begin{figure}
	\centering
	\includegraphics[width=\linewidth]{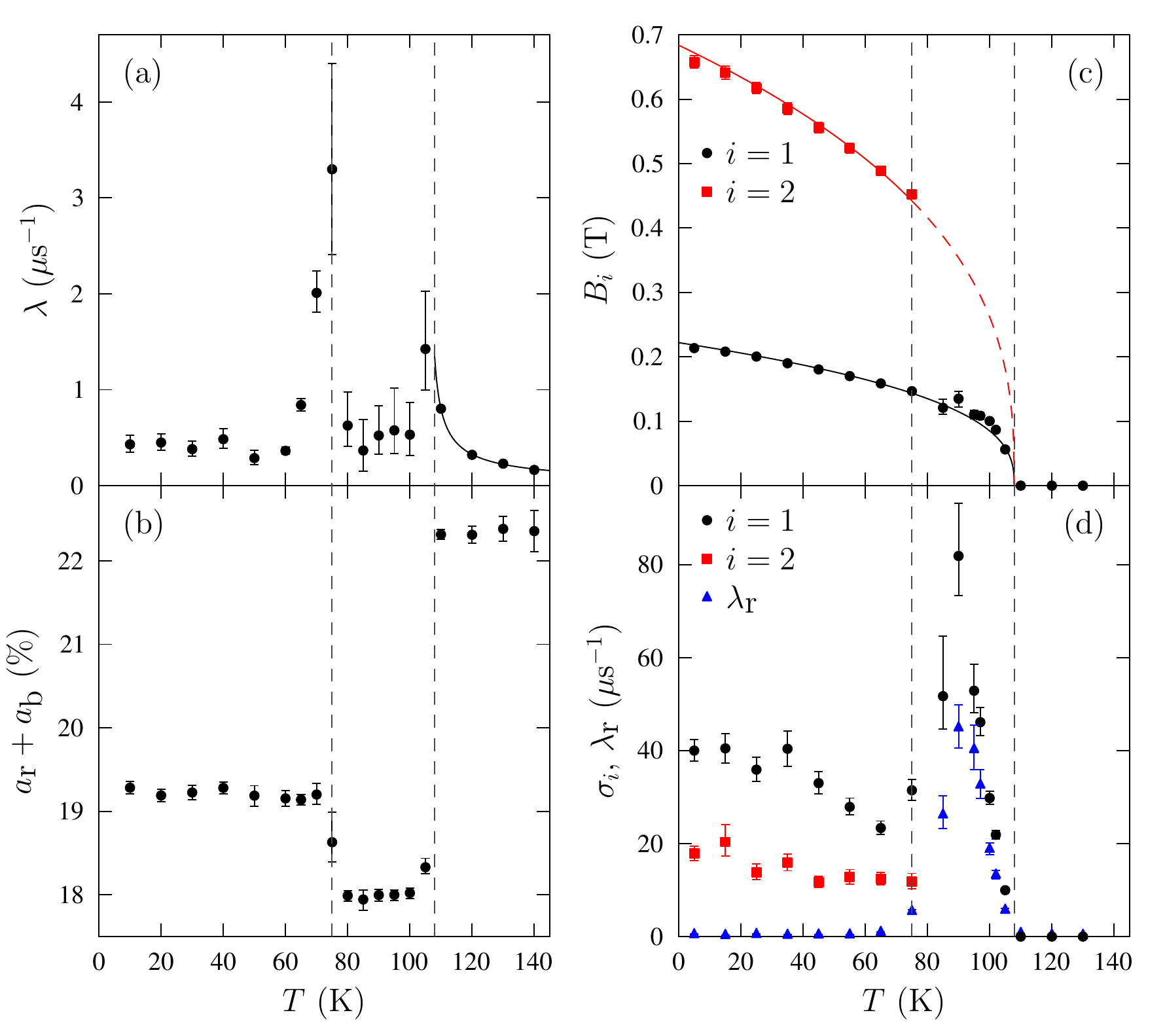}
	\caption{Parameters extracted through fitting ZF $\mu$SR measurements of HgCr$_2$Se$_4$ as described in the text. (a--b) results from ISIS, (c--d) results from S$\mu$S. Dashed lines indicate boundaries between the different regimes as discussed in the text. The solid lines in (a) and (c) are fits to critical behavior typical for a 3D Heisenberg ferromagnet, with the dashed section extending the behavior to $T_{\rm C}$.}
	\label{fig:zf}
\end{figure}

Representative ZF $\mu$SR measurements of HgCr$_2$Se$_4$ are presented in Fig.~\ref{fig:data}, showing both the long-time relaxation [(a), measured at ISIS] predominantly sensitive to dynamics of the system on the muon timescale, and short-time behavior [(b), measured at S$\mu$S], dominated by effects due to long-range magnetic order.
The measurements
performed at ISIS were fitted over the entire temperature range using the function
\begin{equation}\label{eqn:isis}
	A\left(t\right) = a_{\rm r}\exp\left(-\lambda t\right) + a_{\rm b}\exp\left(-\lambda_{\rm b} t\right) ,
\end{equation}
where the component with amplitude $a_{\rm r}$ captures the contribution from muons stopping in the sample
and relaxing with rate $\lambda$ determined in part by dynamic fluctuations of the local magnetic field, while the baseline term with amplitude $a_{\rm b}$ accounts for muons that stop outside of the sample in the silver backing plate and weakly relax with rate $\lambda_{\rm b}$ due to interaction with the Ag nuclear moments.
As expected, the rate $\lambda_{\rm b}=0.01$~$\mu$s$^{-1}$ was found to remain constant at all temperatures.
In a magnetically ordered, polycrystalline sample we expect $1/3$ of muons to stop with their spin initially aligned along the local magnetic field (contributing a slowly relaxing `1/3 tail' to the spectra). 
The other muons, that stop with a component of their spin initially perpendicular to the local magnetic field, will precess and relax, with both effects too rapid to be observed at ISIS due to the pulse-width of the ISIS muon beam.
(Instead, this contribution is seen in the measurements performed at S$\mu$S, discussed below.)
This accounts for a lost fraction of the asymmetry in the ordered state in the ISIS measurements.
Values of these parameters extracted through fitting the data can be seen in Fig.~\ref{fig:zf}(a--b).

Two  peaks in $\lambda$, as well as unusual changes in the total asymmetry $a_{\rm r}+a_{\rm b}$ (typically, the asymmetry remains approximately constant far from $T_{\rm C}$), suggest three distinct regimes of behavior distinguished by their dynamics; we indicate the transition between these regimes with dashed lines in Fig.~\ref{fig:zf}.
At low $T$ there is little relaxation. 
Above $T~\approx~75$~K the drop in $a_{\rm r}+a_{\rm b}$ indicates that an increased number of muons have their spin polarisation relaxed too rapidly to be detected with the time-resolution available at ISIS, suggestive of  dynamics of the local magnetic field on the muon timescale.
This may also suggest that the observed peaks in $\lambda$ are not two distinct peaks, but rather the two edges of a single peak (which will have a maximum between 75~K and $T_{\rm C}\approx 106$~K), with the relaxation rate in this regime too rapid to be detected in these data.
For $T>T_{\rm C}$, the recovery of $a_{\rm r}+a_{\rm b}$ and gradual decrease in $\lambda$ after a peak is suggestive of a transition to a disordered magnetic phase.
Above $T_{\rm C}$, in a 3D Heisenberg magnet (such as HgCr$_2$Se$_4$), we would expect $\lambda\propto\left|T-T_{\rm C}\right|^{-0.709}$~\cite{pelissetto2002critical,pratt2007chiral,pospelov2019non}.
This model describes the data well, with $T_{\rm C}\simeq106$~K, and can be seen as a solid line in Fig.~\ref{fig:zf}(a).

In equivalent measurements performed with higher time resolution at S$\mu$S, oscillations in $A\left(t\right)$ are observed below $T_{\rm C}$ [see Fig.~\ref{fig:data}(b)], indicative of coherent precession of the muon-spin polarisation that occurs due to long range magnetic order.
To characterise these oscillations the data are fitted with the function
\begin{equation}
	\begin{split}
		A\left(t\right) =& \sum_{i=1}^{n}a_{\rm o}^i\cos\left(\gamma_\mu B_i t + \phi_i\right)\exp\left(-\sigma_{i}^2 t^2\right) \\
		&+ a_{\rm r}\exp\left(-\lambda_{\rm r} t\right) + a_{\rm b} ,
\label{eq:smus_fit}
              \end{split}	
\end{equation}
where $a_{\rm o}^i$ the amplitude of the oscillating components, each due to precession in local magnetic field $B_i$ with associated phase offset $\phi_i$ and relaxation rate $\sigma_i$, which is related to the distribution of magnetic fields at that muon site.
The term with amplitude $a_{\rm r}$  again accounts for muons that stop with their spin initially aligned with the local magnetic field direction and hence do not precess, and $a_{\rm b}$ accounts for muons stopping outside of the sample.

At low $T$, below $T~\simeq~75$~K, $n~=~2$ is needed to capture the behavior of $A\left(t\right)$, indicating two magnetically distinct muon stopping sites.
Extracted parameters through fitting of $A\left(t\right)$ are shown in Fig.~\ref{fig:zf}(c--d).
The fields at the muon sites are found to vary in fixed proportion, $B_2/B_1~=~3.08(5)$, with $a_{\rm o}^1 + a_{\rm o}^2 = 9.8(3)$\%, with $a_{\rm o}^2/(a_{\rm o}^1+a_{\rm o}^2)~=~0.281(11)$, $\phi_1~=~7(2)\degree$ and $\phi_2~=~-45(3)\degree$ found to be independent of $T$ over this range.
All these factors suggest that the two distinct magnetic environments occur due to the same magnetic order parameter, typical of ferromagnetic behavior previously observed at low $T$ in HgCr$_2$Se$_4$~\cite{Lin2016}.
The non-zero (and distinct) phase offsets of the two components are unusual, and suggest complex magnetic behavior that leads to differences in the local magnetic field distribution at the two muon stopping sites. Specifically, the $\phi\approx -45^{\circ}$ phase offset is sometimes observed for the case of an incommensurate magnetic structure \cite{blundell1999spin, blundell2021}. (We return to this feature below.)
The rate $\lambda_{\rm r}$ matches that found in the ISIS measurements discussed above.

The relaxation rates $\sigma_{i}$ of the oscillating components in a static, magnetically-ordered phase can be related to the width of the distribution of magnetic fields at the muon site via $\sigma_{i}^2~=~\Delta_{i}^2/2$, where $\Delta_{i}~=~\gamma_\mu\sqrt{\left<(B_i-\langle B_{i} \rangle)^2\right>}$.
We observe  $\sigma_2<\sigma_1$, which is unexpected as we  would typically expect $\sigma_{i}$ to scale with the field at the muon site.
This further suggests that the two distinct muon stopping sites
have different static local environments and/or different dynamics.

As $T$ is increased above 75~K, oscillations in $A\left(t\right)$ are still observable, but are heavily damped [see Fig.~\ref{fig:data}(b)].
This indicates the system still exhibits long-range magnetic order, but there is a much greater dynamic response on the muon timescale (as is also observed in the measurements performed at ISIS).
Only $n=1$ is needed to parametrize these data in this regime. However, the large relaxation rates $\sigma_{1}$, coupled with the smaller relative amplitude seen at low $T$, suggests that the contribution to $A\left(t\right)$ from $B_2$ is relaxed too quickly to be observed.
This is consistent with the increased lost asymmetry observed in Fig.~\ref{fig:zf}(b).
In this regime $\phi_1~=~7(3)\degree$ (consistent with $\phi_1$ at low $T$) and $a_{\rm r}/(a_{\rm o}^1+a_{\rm r})~=~0.391(15)$.
As expected from the ISIS data, $\sigma_1$ and $\lambda_{\rm r}$ both peak in the middle of this temperature regime. Although it is not normally possible to distinguish an increase in $\sigma_1$ occuring due to an increase in the width of the distribution of magnetic fields at the muon site, or a change in the dynamic fluctuation rate on the muon timescale, we infer from the behavior of lambda in the ISIS data that this effect is likely dominated by dynamic effects. This further suggests the peaks observed in $\lambda$ in Fig.~\ref{fig:zf}(a) are the edges of a single peak.
The peak in $\sigma_{1}$ might suggest that relaxation pathways freeze out, reducing in frequency and passing through the frequency window to which $\mu$SR is sensitive on cooling below 90\,K, as the behavior changes due to the system gradually changing state.

For $T~<~T_{\rm C}$ we expect critical behavior of $B_i$ such that $B_i~=~B_i^0\left(1-T/T_{\rm C}\right)^\beta$, with Ref.~\cite{Lin2016} identifying $\beta~=~0.361$, close to the theoretical value of $\beta~=~0.367$ for a 3D Heisenberg ferromagnet.
Due to the heavy suppression of the oscillatory part of $A\left(t\right)$ just below $T_{\rm C}$, extracting a reliable value of $\beta$ is challenging. However constraining to $\beta~=~0.367$ gives a good fit of the extracted $B_{\rm i}$, as seen in Fig.~\ref{fig:zf}(c), consistent with the system behaving as a 3D Heisenberg ferromagnet,  giving a critical temperature $T_{\rm C}~=~107.9(3)$~K.
At high $T$ there are no oscillations observable in $A\left(t\right)$, consistent with a lack of long-range magnetic order.
Eqn.~\ref{eqn:isis}, with $\lambda_{\rm b}~=~0$, describes the behavior in this regime.

Finally, we performed weak transverse field (wTF) measurements in order to check for evidence of magnetic phase separation, which are detailed in Appendix~\ref{sec:wtfmusr}. These measurements do not show evidence for such phase separation down to the few-percent level, but do show similar transitions to those reported from the ZF measurements. 
The key observation from these $\mu$SR results is the spin dynamics seen in the intermediate temperature regime. In order to link these to charge-carrier dynamics, we next turn to the results of fluctuation spectroscopy measurements.


\subsection{Fluctuation (noise) spectroscopy} 
\paragraph*{Sample characterization -- DC conductivity}
Transport measurements have been carried out on several samples of $n$-type HgCr$_2$Se$_4$ single crystals. From the literature it is known that 
the transport properties and the CMR effect are highly sensitive to details in the growth procedure and sample treatment \cite{Solin2008}. However, the magnetism and basic transport characteristics are consistent for the different types of samples. For the time-resolved transport measurements presented here, we chose a single-crystalline sample with an electron density of $n \approx 6.1 \times 10^{15} {\rm cm}^{-3}$ at liquid helium temperatures, the impedance range of which is convenient for standard fluctuation spectroscopy measurements.
\begin{figure}[t]
\centering
\includegraphics[width=0.85\columnwidth]{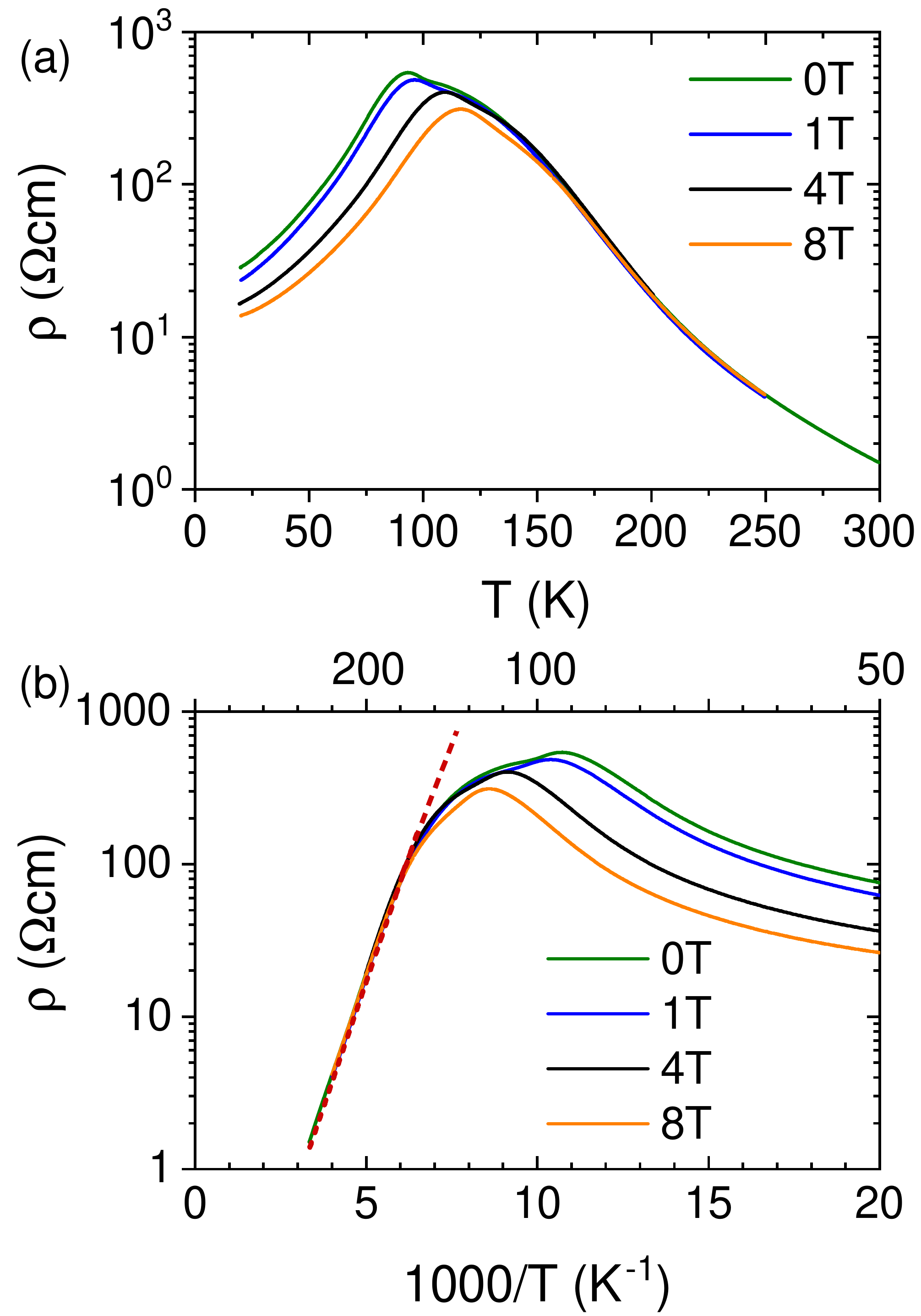}
\caption{(a) Longitudinal resistivity $\rho_{xx}(T)$ of $n$-type HgCr$_2$Se$_4$ for different externally applied magnetic fields of $B$ = 0, 1, 4 and 8\,T. (b) Arrhenius plot $\log{\rho_{xx}}$ vs.\ $1/T$ for magnetic fields $B$ = 0, 1, 4 and 8\,T. The linear fit (red dashed line) (see text) indicates a thermal activation gap of $\Delta_b = 133$\,meV.} 
\label{resistivity}%
\end{figure}%
For the present sample, due to a higher carrier mobility, the absolute values of the resistivity are considerably lower than for the sample discussed in our earlier work \cite{Lin2016} and hence the CMR effect less pronounced. However, 
the transport properties of the HgCr$_2$Se$_4$ sample shown in Fig.\ \ref{resistivity} exhibit the same general characteristics, 
namely a semiconducting behavior upon cooling from room temperature followed by an intermediate temperature regime, where the resistivity increases less strongly, and a drop and metallic behavior at $T \lesssim 92$\,K, which is somewhat below the ferromagnetic ordering temperature $T_C \approx 107$\,K determined from $\mu$SR or magnetic susceptibility measurements (see Appendix~\ref{Appendix_magnetic}), but is close to the relaxation rate peak seen in the ZF $\mu$SR. Similar to the sample in \cite{Lin2016}, the resistivity in the intermediate and metallic regimes, is suppressed by magnetic fields. 
The Arrhenius plot in Fig. \ref{resistivity}(b) shows a thermally activated behavior $\rho(T) = \rho_0 \exp{(\Delta_b/k_BT)}$ of the resistivity at high temperatures with an activation energy $\Delta_b = 133$\,meV.\\

\begin{figure}[h!]
\centering
\includegraphics[width=0.9\columnwidth]{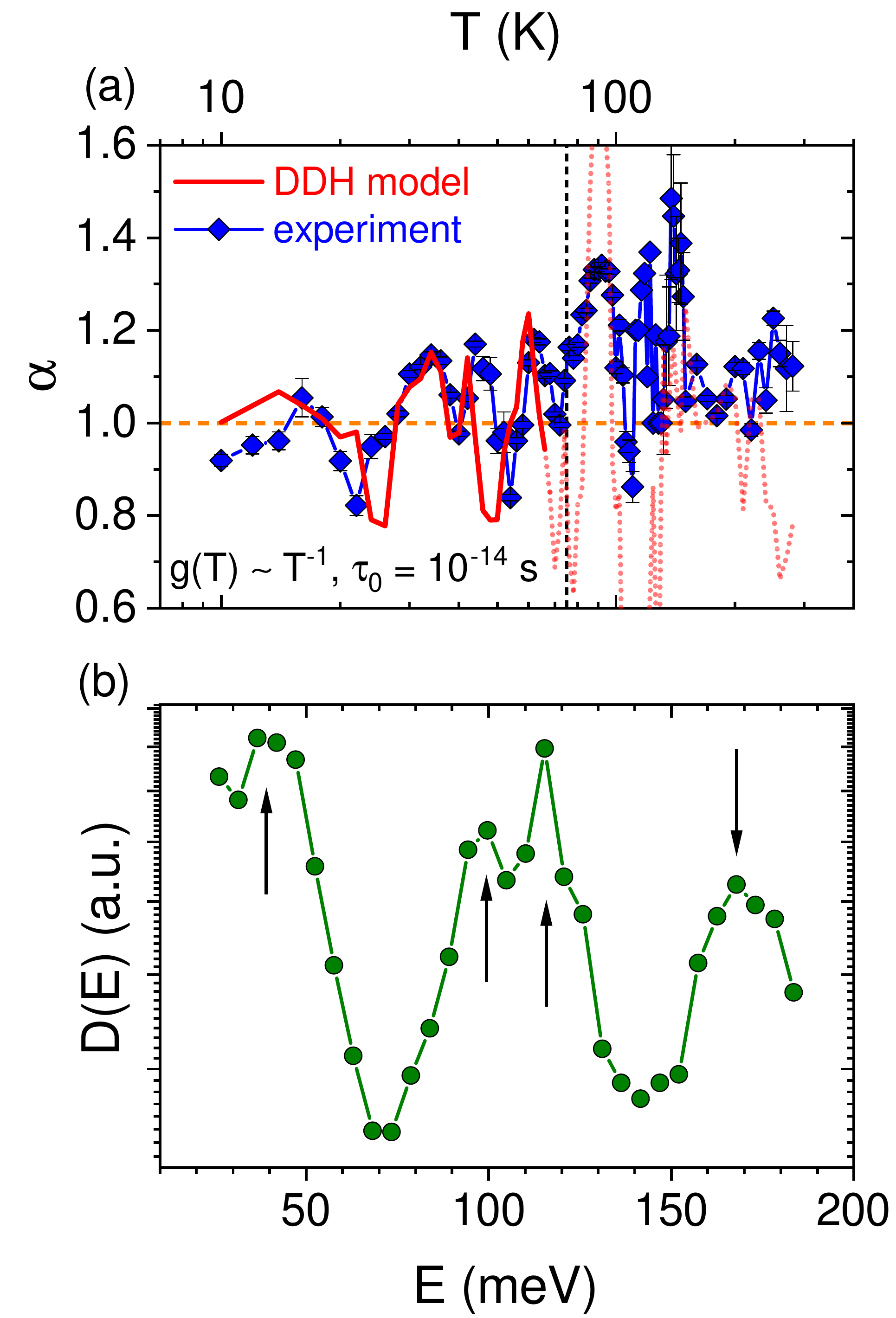}
\caption{(a) Temperature dependence of the frequency exponent $ \alpha$ compared to the values calculated from the DDH model (red line), Eq.\ (\ref{alpha}). The parameters for $g(T)$ and $\tau_0$ are given. Dashed line marks the temperature below which the DDH model is valid. 
(b) Distribution of activation energies $D(E)$ calculated from Eq.\ (\ref{DvonE}), which determines the temperature dependence of the noise magnitude $S_R/R^2(T)$ at $f = 1$\,Hz below $T = 75$\,K, see text.} 
\label{DDH}%
\end{figure}

\paragraph*{Magnetic polaron dynamics and percolation (\,$T > T_C$)} 
\begin{figure}[b]
\centering
\includegraphics[width=\columnwidth]{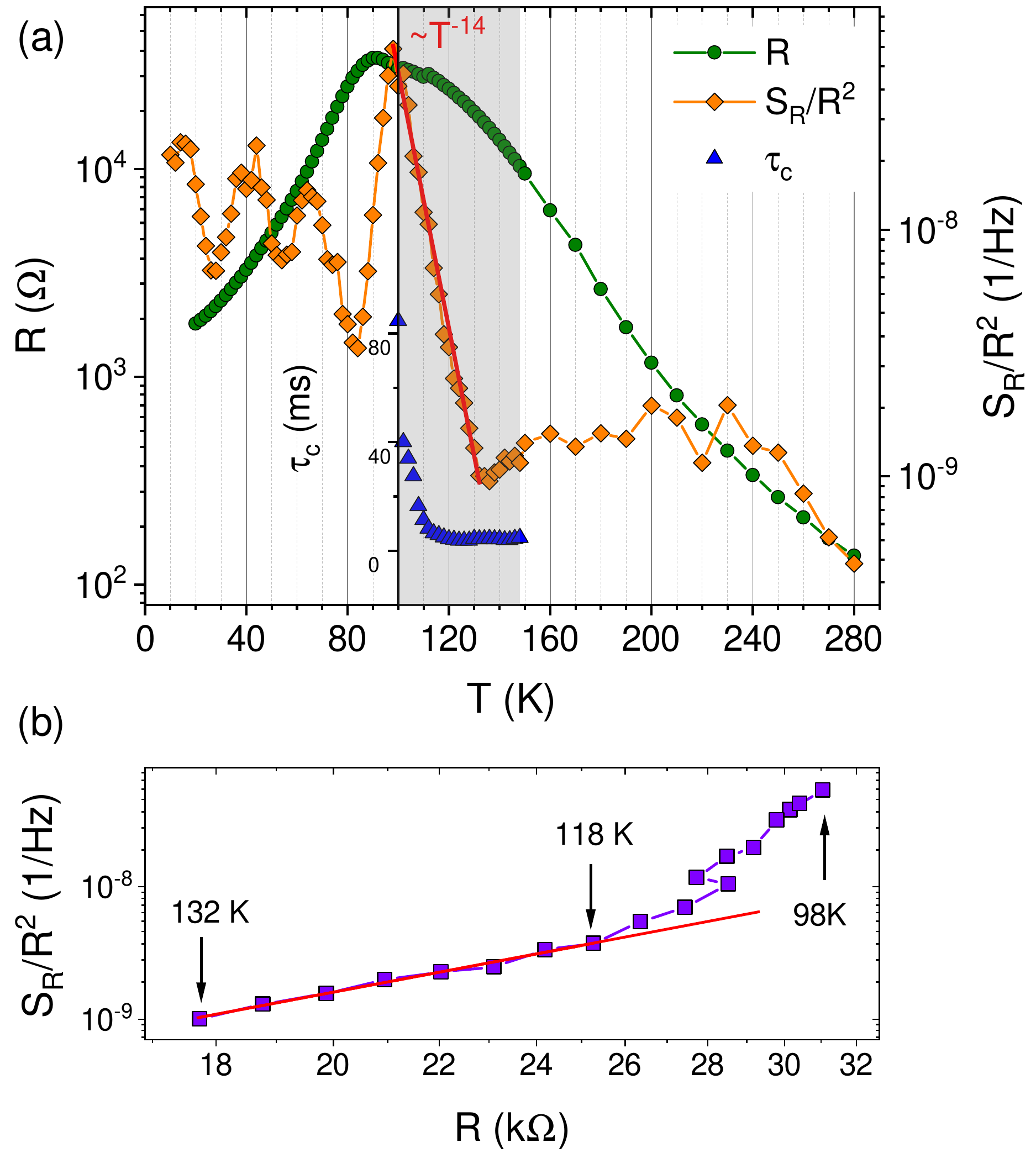}
\caption{(a) Resistance, $R$, (green symbols) and normalized resistance noise PSD, $S_R/R^2 (f = 1\,{\rm Hz})$, (orange) of the $1/f$-type noise vs.\ temperature $T$. In the grey shaded region additional Lorentzian-type spectra due to two-level switching processes were observed. Their amplitude is {\it not} included here. Their characteristic time constant $\tau_c$, however, is shown as blue triangles.
 In the percolation regime (see text), the noise PSD strongly increases with decreasing temperature following an approximate power law $S_R/R^2 \propto T^{-14}$ (red line). 
(b) Scaling $S_R/R^2 \propto R^w$ in the percolation regime with $w = 3.6 \pm 0.1$ (red line). Below about 120\,K the scaling exponent becomes temperature dependent.} 
\label{noise}
\end{figure}
Next, we discuss the overall behavior of the resistance fluctuations in zero magnetic field. Figure \ref{noise}(a) displays the sample resistance, $R$, at discrete temperatures (green circles) and the corresponding normalized resistance noise PSD, $S_R/R^2$, (orange diamonds) of the $1/f$-type noise taken at $f = 1$\,Hz. In our frequency range $f = 0.01 - 100$\,Hz, we observe pure $S_R/R^2 \propto1/f^\alpha$ noise for temperatures $T > 150$\,K and $T < 100$\,K, with a frequency exponent $\alpha(T)$
exhibiting values between 0.8 and 1.4, see Fig.\ \ref{DDH}(a) below, where a variation from 1 to larger or smaller values implies a shift of the spectral weight (and therefore the fluctuation dynamics) to lower or higher frequencies, respectively, as compared to the homogeneous distribution of time scales for $\alpha = 1$ \cite{Mueller2018}, see also Appendix~\ref{Appendix_noise}. In contrast, a Lorentzian spectrum characteristic for distinct two-level fluctuations is superimposed on the $1/f^\alpha$-noise for $100\,{\rm K} < T < 150\,{\rm K}$ (grey-shaded area in Fig.\ \ref{noise} (a)). Therefore, 
the observed spectra can be described by
\begin{equation}
\frac{S_R(f,T)}{R^2} = \frac{A_1}{f^\alpha} + \frac{A_2 f_c}{f^2 + f_c^2},
\label{noise_spectra}
\end{equation}
where $A_1(T)$ denotes the magnitude of the $1/f$-type noise and $\alpha(T)$ its frequency exponent. The magnitude of the Lorentz contribution $A_2(T) \equiv 2/\pi \cdot \langle (\delta R)^2 \rangle/R^2$ and $f_c \equiv 1/(2 \pi \tau_c)$ denote the relative change of the global resistance due to the local fluctuation and the corner frequency, respectively, see Fig.\ \ref{spectra}(b). The action of a single fluctuator giving rise to the Lorentz spectrum (and $A_2(T) \neq 0$) becomes enhanced in our frequency range ('noise window' $[0.01\,{\rm Hz}, 100\,{\rm Hz}]$) only for $100\,{\rm K} < T < 150\,{\rm K}$. For $T > 150$\,K and $T < 100$\,K, however, the magnitudes $A_2(T) = 0$ and $A_1(T) \equiv S_R/R^2(f = 1\,{\rm Hz},T)$ as well as the frequency exponent $\alpha(T) \equiv (-\partial \ln{S_R/R^2(f,T)})/(\partial \ln{f})$ can be directly read from the fit shown exemplarily for a selected temperature in Fig.\ \ref{spectra}(a). 

In Figure \ref{noise}(a), for all temperatures including the grey-shaded area, we show   
$A_1(T)$ (orange diamonds), i.e.\ the magnitude of the Lorentzian contribution $A_2(T)$ is {\it not} included [see Appendix~\ref{Appendix_noise} for the behavior of $A_2(T)$]. Clearly, the resistance fluctuations $S_R/R^2(T)$ behave very differently than the time-averaged mean $R(T)$ and reveal information on the low-frequency dynamics of the charge carriers coupled to the magnetic (and possibly other) degrees of freedom. Upon cooling from room temperature, the $1/f$-noise initially slightly increases until below $\sim 230 - 240$\,K it stays roughly constant down to about 150\,K (roughly coinciding with the onset of two-level fluctuations), below which it again slightly decreases. At about 135\,K the $1/f$-noise then shows a sudden increase of almost two orders of magnitude, peaks at around $T_{\rm max} \approx 98$\,K (slightly below $T_C$ and above the resistivity maximum) and exhibits a sudden drop upon further cooling through the magnetically ordered phase. Below a sharp minimum, the noise then increases again and shows additional local maxima and minima upon further cooling. As will be described below, these features exhibit a qualitatively different dynamics than the sharp peak at 98\,K and their behavior can be well understood within a model of non-exponential kinetics, i.e.\ a superposition of many independent two-level fluctuators with a characteristic distribution of thermal activation energies. 
We interpret the strong increase of the $1/f$-noise below about $135$\,K following an approximate power law $S_R/R^2 \propto T^{-14}$, the sharp 
peak around 98\,K and the even more pronounced drop on the low temperature side as a percolation transition of correlated magnetic polarons quite similar to the observations in other CMR systems, as e.g.\ in perovskite manganites \cite{Podzorov2000} and EuB$_6$ \cite{Das2012}. (Notably, a strong, power-law increase of the fluctuations is a general feature observed upon approaching a metal-to-insulator transition \cite{Bogdanovic2002,Kar2003,Hartmann2015}.) As expected for a simple random resistor network \cite{Kogan1996}, we observe a scaling behavior, see Fig.\ \ref{noise}(b), demonstrating a power-law $S_R/R^2 \propto R^w$ 
upon approaching the percolation threshold, where $w = \kappa/t$ depends on the percolation scenario \cite{Tremblay1985,Kogan1996}. 
We find $w = 3.6 \pm 0.1$, albeit only in the small temperature region of the strong increase of the 1/$f$-noise. Below this temperature regime the scaling exponent becomes temperature dependent, which may be due to the strong influence of spin correlations on the effective size of magnetic polarons to be discussed below. Large values of $w \sim 3$ have been found, e.g., in random metal-insulator composites, where a modified random void model is invoked, in which the stochastic voids in a conducting matrix are replaced by interpenetrating conducting spheres in an insulating matrix \cite{Rudman1985}. 

\begin{figure}[t]
\centering
\includegraphics[width=0.9\columnwidth]{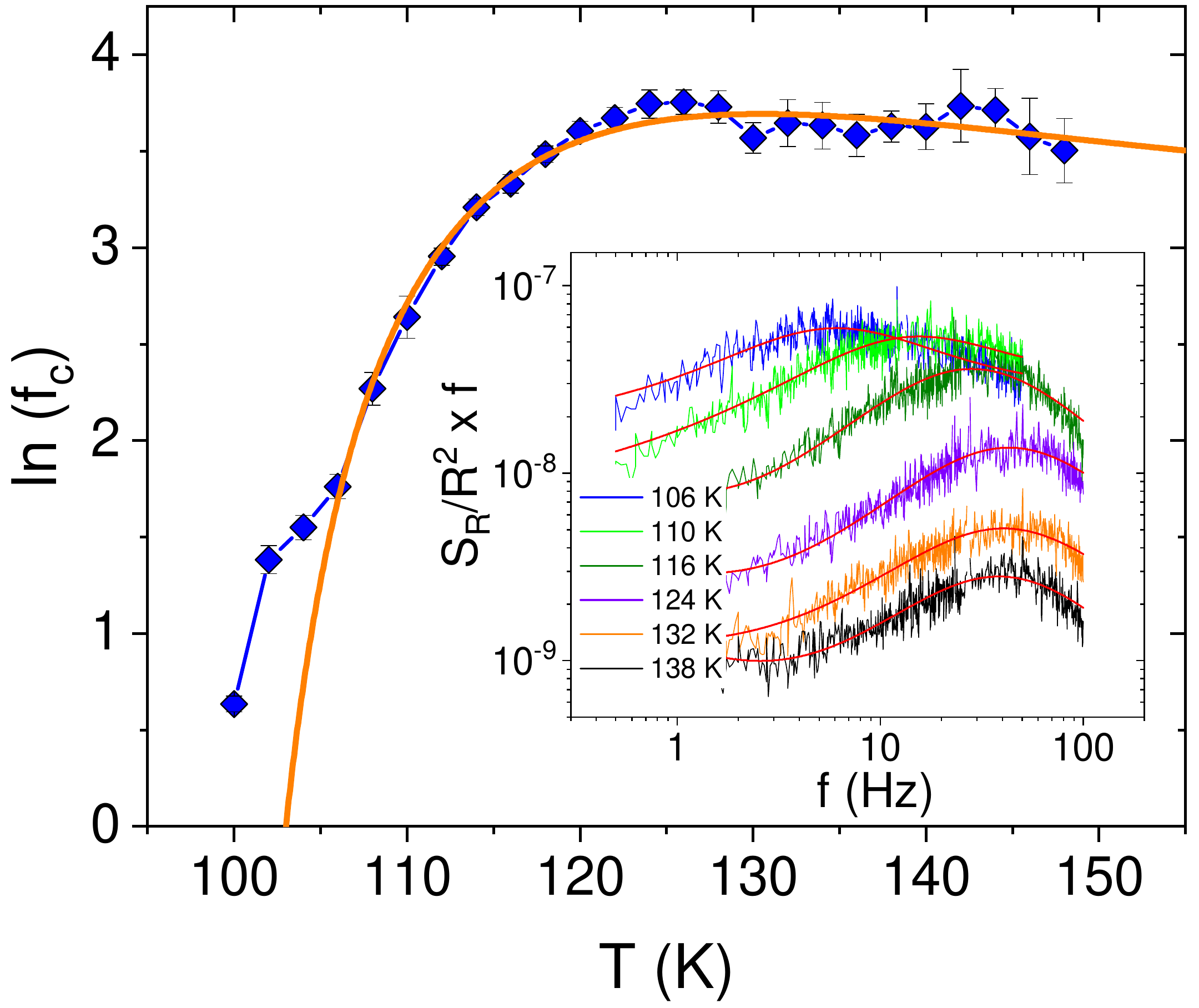}
\caption{Temperature dependence of the corner frequency $f_c$ of the two-level fluctuations observed for $100\,{\rm K} < T < 150\,{\rm K}$ (grey-shaded area in Fig.\ \ref{noise} (a)) in zero magnetic field shown as $\ln{f_c}$ vs.\ $T$. Orange line is a fit to the model Eq.\ (\ref{model}), see text and \cite{comment3}. The inset shows corresponding spectra at selected temperatures with fits (red lines) to Eq.\ (\ref{noise_spectra}).} 
\label{corner-frequency}
\end{figure}

The temperatures $T^\ast \sim 140 - 150$\,K, where the magnetic susceptibility of our sample deviates from a Curie-Weiss law, see Appendix~\ref{Appendix_magnetic}, may be interpreted as the crossover from isolated polarons to a percolative behavior \cite{Lin2016}. In electron spin resonance measurements \cite{Li2018} this temperature scale is assigned to the occurrence of an additional magnetic contribution associated to magnetic correlations among MPs or ferromagnetic clusters.\\
A clue for better understanding the slow dynamics within this percolation regime 
comes from the distinct two-level fluctuations being enhanced in our 'noise window' for $100\,{\rm K} < T < 150\,{\rm K}$ (see above). Figure \ref{noise}(a) displays in this region (grey-shaded area) the time constant $\tau_c = 1/(2 \pi f_c)$ of the two-level fluctuator as blue triangles, determined from the fits of the spectra to Eq.\ (\ref{noise_spectra}) [see Fig.\ \ref{spectra}(b) and inset of Fig.\ \ref{corner-frequency}]. We find that for decreasing temperatures the characteristic relaxation time $\tau_c$ is nearly constant before below about 120\,K --- the same temperature where the percolation scaling shown in Fig.\ \ref{noise}(b) breaks down --- it strongly increases corresponding to a drastic slowing down of the fluctuation dynamics upon approaching the magnetic transition 
and 
the $1/f^\alpha$-noise maximum. 

This unusual behavior is also highlighted in Fig.\ \ref{corner-frequency} showing the corresponding temperature dependence of the corner frequencies $f_c$ plotted as $\ln{f_c}$ vs.\ $T$ with the evolution of the spectra fitted by Eq.\ (\ref{noise_spectra}) shown in the inset. Clearly, the behavior expected for a simple thermally activated process $f_c = f_0 \exp{(-E_a/k_BT)}$ is {\it not} observed. Instead, $f_c$ stays roughly constant upon cooling from 150\,K to about 120\,K below which it gradually more strongly decreases, i.e.\ the characteristic time of the fluctuations slows down, with an increasingly steeper slope (stronger slowing down) upon lowering the temperature. A natural explanation for a two-level-fluctuator in a system with magnetically-induced electronic phase separation and percolation 
is the switching of a bound magnetic polaron, i.e.\ a nano-scale magnetic cluster with aligned spins having to overcome an energy barrier due to the environment of the surrounding paramagnetic matrix. 
The logarithmic growth of the radius of the bound magnetic polaron with decreasing temperature described in the seminal works by Bhatt et al.\ \cite{Bhatt2002} and Kaminski and Das Sarma \cite{Kaminski2002} indeed suggests a slowing down of the polaron dynamics until the percolation threshold is reached, where the MPs merge and the impedance of their motion in the insulating matrix becomes lifted. 
However, we will show in the following that in such a picture the observed behavior can only be understood when the MPs effective radius depends on the spin correlation length, i.e.\ a core-shell structure of the MP is considered. Indeed, such spin correlations have been proposed to play an important role for a wide temperature range $T_C < T < T^\ast$  in the paramagnetic phase \cite{Lin2016}.

In the model of correlated polarons by Bhatt et al.\ and Kaminski and Das Sarma, for temperatures smaller than a characteristic energy scale given by the localization length of a dopant (captured hole), all magnetic spins within distance $r_p(T) = \frac{1}{2} a_B \ln{(s S |J_0|/k_B T)}$ of dopants align antiferromagnetically, where $a_B$ is the localization length of the donor, $s$ the carrier spin, $S$ the spin of the magnetic ion in the matrix and $J_0$ the effective exchange constant \cite{Bhatt2002, Kaminski2002}. We now consider a core-shell structure, where $r_{core} = r_p(T)$ describes the radius over which the magnetic moments are bound to those of the carrier and $r_{shell} = \xi_0(T/T_C - 1)^{-\nu}$ the decay of the spin polarization at the MP boundary determined by the spin correlation length, with $\nu = 0.709$ for a 3D Heisenberg ferromagnet.  Therefore, $r_{MP}(T) = r_{core}(T) + r_{shell}(T)$ and for the MP's volume 
\begin{equation}
V_{MP}(T) \approx 4 \pi r_p(T)^2 \left[ \frac{1}{3}r_p(T) + \xi_0 \left(\frac{T}{T_C} -1\right)^{-\nu} \right].
\end{equation}
Assuming a bistable switching as suggested by the experimental observation and a thermally activated behavior $f_c(T) = f_0 \exp{(-KV_{MP}/k_BT)}$ with a parameter $K$ determining the energy barrier, we have
\begin{widetext}
\begin{equation}
\ln{f_c} \approx \ln{f_0} - \frac{\pi a_B^2 K}{k_BT}\left( \ln{\frac{s S |J_0|}{k_B T}} \right)^2 \cdot \left[\frac{a_B}{6}\ln{\frac{s S |J_0|}{k_B T}} + \xi_0 \left( \frac{T}{T_C} -1 \right)^{-\nu} \right].
\label{model}
\end{equation}
\end{widetext}
The orange line in Fig.\ \ref{corner-frequency} is a fit to the model Eq.\ (\ref{model}) with a set of parameters with reasonable values \cite{comment3} demonstrating that only when considering the significant spin correlations in HgCr$_2$Se$_4$ \cite{Lin2016} it is possible 
to reproduce the highly unusual temperature dependence of the corner frequency of the observed two-level switching. We interpret this behavior in a sense that the exceptionally strong increase of the $1/f$-type noise upon cooling is due to two effects, namely (i) the percolation of MPs in a complex network and (ii) their increasingly slow `internal' dynamics. This is also reflected in the distribution of the spectral weight of the $1/f$-noise, see Appendix~\ref{Appendix_noise}. 
Although 
the fit is not unique due to the large set of independent parameters \cite{comment3} we consider the good agreement as a signature of the particular importance of spin correlation effects for the MP percolation in HgCr$_2$Se$_4$. Interestingly, in the related spinel CdCr$_2$S$_4$, a spin-pair correlation driven CMR effect is discussed \cite{Xia2019} which indicates a more general importance of our findings. As an independent cross check, we find that the estimated percolation threshold $r_{\rm perc} \approx 0.86/n^{1/3}$ with $n$ the density of MP \cite{Kaminski2002} yields a radius of the polarons which agrees quite well with the values estimated from our model, Eq.\ (\ref{model}), at the 'percolation temperature', i.e.\ the $1/f$ noise maximum at 98\,K.\\

\paragraph*{Non-exponential kinetics (\,$T < T_C$)} 
Now we turn to the charge fluctuations in the magnetically ordered state. A widely used model for the ubiquitous $1/f^\alpha$-noise in solids is a superposition of independent `fluctuators', i.e.\ two-level processes fluctuating with a characteristic time constant $\tau$, which couple to the resistance of the system, each contributing a Lorentzian spectrum \cite{Kogan1996,Raquet2001,Mueller2018}: 
\begin{equation}
\frac{S_R(f,T)}{R^2} \propto \int\limits_0^\infty D(\tau) \frac{4\tau}{1+4 \pi^2 f^2 \tau^2}{\rm d}\tau,
\end{equation}
where $D(\tau)$ is a distribution of effective time constants.
While in our case such a model is not expected to be valid in the percolation regime above and around $T_C$ or $T_{\rm max}$, its applicability at lower temperatures can be tested. The model assumes a thermally-activated switching processes where the relaxation time follows an Arrhenius law
$\tau = \tau_0 \exp{(E_a/k_BT)}$, 
with $\tau_0$ an attempt time, typically of order of the inverse phonon frequencies, and $E_a$ the a characteristic energy of the two-level potential. The superposition of Lorentzian spectra with a sufficiently broad distribution of activation energies $D(E_a) \equiv D(E)$ then leads to a $1/f^\alpha$-type spectrum \cite{Pre1950} in a frequency window determined by the energy distribution. A phenomenological description based on this concept of non-exponential kinetics \cite{comment2} has been introduced by Dutta, Dimon and Horn (DDH) \cite{Dutta1979}. Under the assumption that the same fluctuation processes determine the temperature dependences of both the noise magnitude [see Fig.\,\ref{noise}(a)] and the frequency exponent (spectral weight distribution)  [see Fig.\,\ref{DDH}(a)], the model predicts
\begin{equation}
\alpha(T) = 1 - \frac{1}{{\rm ln}(2 \pi f \tau_0)} \left[ \frac{\partial {\rm ln}S(T)}{\partial {\rm ln} (T)} - \frac{\partial {\rm ln}g(T)}{\partial {\rm ln} (T)} -1 \right],
\label{alpha}
\end{equation} 
where $g(T)$ describes the coupling of the two-level processes to the resistance and accounts for an explicit temperature dependence of the distribution of activation energies \cite{Raquet1999}.
For $g(T) \propto T^{-1}$, creating a constant offset of the model adaption in Fig.\ \ref{DDH}(a) and a typical inverse phonon frequency $\tau_0 = 10^{-14}$\,s, the prediction Eq.\,(\ref{alpha}) matches the data quite well below about 75\,K (i.e.\ the boundary of the intermediate temperature regime identified via our $\mu$SR measurements) but deviates considerably above that temperature, in particular where the noise is mainly caused by the percolation of MPs. Therefore, for energies $E < k_B T \ln{(2 \pi f \tau_0)^{-1}}$ with $T = 75$\,K and at $f = 1$\,Hz, one can determine the energy distribution causing the observed $1/f^\alpha$-noise \cite{Dutta1979} from
\begin{equation}
D(E) \propto \frac{2 \pi f}{k_BT}\frac{S_R(f,T)/R^2}{g(T)},
\label{DvonE}
\end{equation}
which is shown in Fig.\,\ref{DDH}(b). Clearly, the observed magnitude of the $1/f$-type noise and its distribution of spectral weight can be described by a series of local maxima/peaks in the energy distribution $D(E)$ of thermally-activated fluctuators at about 36, 100, 115 and 168\,meV, see arrows in Fig.\ \ref{DDH}(b).
It is inherent to the model that the two-level fluctuators are not specified {\it a priori} but may be identified {\it a posteriori} from the maxima in $D(E)$, i.e.\ their energy signature. As discussed below, based on the $\mu$SR results we attribute these energies to clusters of MPs that form around the magnetic and percolation transition, become frozen at low temperatures and fluctuate as magnetic domains or dynamic domain walls. 


\section{Discussion} 
Naturally, resistance noise and $\mu$SR spectroscopy probe different frequency regimes.
The latter is a local probe of the microscopic  magnetic field distribution, whose dynamic response is determined by $\gamma_{\mu}B$ where $B$ is the magnitude of the total field at the muon site, such that $\mu$SR typically probe spectral features in the  MHz -- GHz range such as low-energy spin fluctuations or spin diffusion.
In contrast, resistance noise spectroscopy is sensitive to the dynamics of collective objects, such as magnetic polarons but also nano- or micro-scale magnetic domain walls or domains in ferromagnets.
Our data obtained by the combination of these two methods can be reconciled with a description of magnetism in HgCr$_2$Se$_4$ given in terms of MPs, in which dilute polarons form for temperatures $T > T_C$ and then coalesce below $T_C$ until an ordered magnetic state is stabilized at temperatures far below $T_C$.

The noise measurements reveal low-frequency dynamics consistent with a percolation transition of MP, where there is a strong increase and power-law divergence of the magnitude of the $1/f$-type fluctuations upon approaching the percolation threshold, very similar to the behavior of EuB$_6$ \cite{Das2012}, where the existence of MPs explaining the CMR effect is well established \cite{Suellow2000,brooks2004magnetic,Urbano2004,Zhang2009}. In the present spinel system, however, the density of isolated MPs seems to be considerably lower 
which results in the enhancement of distinct two-level fluctuations superimposed on the underlying $1/f$ noise. We have shown that the MP dynamics can be described by a core-shell model corroborating the importance of spin correlations suggested previously.
The $\mu$SR data suggest that these isolated MPs do not significantly affect the dynamics of the bulk system on the $\mu$SR timescale, or they do not have sufficient density to lead to relaxation of the muon-spin. A rough estimate of the volume fraction occupied by MPs, assuming a radius of $\approx 5$~nm and a density equal to the carrier density $n$, gives 0.3\%, well below the volume at which we would expect $\mu$SR to be sensitive, which is typically of order of a few percent.

This picture is consistent with the noise data, as whilst the isolated MPs do not form much of the total volume fraction of the sample (as seen here with $\mu$SR), they can still have profound effects on the transport properties.
We mentioned that in EuB$_6$~\cite{fisk1979magnetic}, both resistance noise and $\mu$SR spectroscopy provided evidence for the coalescing of magnetic polarons~\cite{brooks2004magnetic} in a similar manner to the scenario we invoke here to explain the behavior of HgCr$_2$Se$_4$. In EuB$_6$, changes in the different parts of the $\mu$SR signal indicated two magnetic transitions, along with a phase separation, above $T_C$, into ferromagnetically ordered regions of overlapping MPs and paramagnetic regions between them. Some of the changes we observe in the $\mu$SR signal measured in the intermediate temperature regime (75\,K $<$ $T$ $<$ $T_C$) of HgCr$_2$Se$_4$ are broadly similar to those found in EuB$_6$ at temperatures above $T_C$ but below its higher-temperature transition \cite{brooks2004magnetic}. However, we do not see evidence for phase separation into paramagnetic and ordered regions in the intermediate temperature range of HgCr$_2$Se$_4$, nor a clear signal in the $\mu$SR above $T_C$ for MP dynamics or freezing.

An interesting feature is the coupling of the magnetism and electronic behavior, specifically the magnetic-order-driven metal-insulator transition, and the coincidence of peaks in $\mu$SR relaxation rates and electronic noise spectroscopy measurements. The magnetism at these temperatures is dynamic on the $\mu$SR time scale as evidenced by the peak in relaxation at $\sim 95$\,K which, remarkably, coincides with a peak both in the electrical resistivity and the normalised resistance fluctuations of the sample, suggesting a degree of magnetoelectric coupling. Further evidence for this coupling comes from the magnetic-order-driven metal-insulator transition. As the normalised PSD is measured at much lower frequencies ($\lesssim$ 100 \,Hz) than those to which $\mu$SR is sensitive, it suggests those dynamics persist over a wide range of frequencies. This holds true for frequencies down to the mHz regime which shows that the relaxation pathways freezing out consist of spins correlated regions over a wide range of length scale up to nano- and micro-scale objects, as e.g.\ clusters of MPs or magnetic domains, which naturally explains the enhanced $1/f$-type noise below $T_C$.

Upon further cooling, a change in dynamics is again observed at different time scales in both spectroscopy methods. Below $\simeq 75$\,K there is little dynamic relaxation at $\mu$SR time scales and the two different local environments (muon stopping sites) are observed to show distinct behavior. This can be interpreted in terms of 
one of the two precession frequencies ($i=1$ in Eq.~\ref{eq:smus_fit})  arising from muons that stop inside a polaron or a cluster of polarons, and the other ($i=2$, with its phase offset indicative of a complicated, possibly incommensurate, field distribution) from muons that stop in between, in domain-wall-like regions. These two muon stopping sites couple to the same magnetic order parameter, but experience different dynamics. Likewise, below the same temperature of about 75\,K the observed $1/f$-type resistance fluctuations are well described by a model of non-exponential kinetics (which inherently implies an `inhomogeneous' distribution of relaxation times $D(E)$ of width much larger than the thermal energy), i.e.\ a change in the charge carrier dynamics coupled to the magnetism occurs. The observed dynamics in $\mu$SR and resistance noise may be consistently explained by large, polarized and domain-wall-like regions remaining from the freezing of the spin dynamics at the MP percolation transition below $T_C$. 
Local ferromagnetic inhomogeneities, which were attributed to frozen clusters of MPs including a large connected (or `infinite') cluster of linked MP forming below $T_C$ indeed have been suggested for the model CMR system EuB$_6$ from measurements of the local magnetic induction \cite{Pohlit2018}.
The low-frequency fluctuation processes in HgCr$_2$Se$_4$ with energies between about 30 and 170\,meV, see Fig.\ \ref{DDH}(b), then are related to thermally-activated switching processes of magnetic domains or the motions of domain walls. 

\section{Conclusion}
The combination of $\mu$SR and resistance fluctuation spectroscopy measurements of the ferromagnetic spinel HgCr$_2$Se$_4$ allows the study of magnetic and charge dynamics over a wide range of frequencies, revealing a degree of magnetoelectric coupling and distinct regimes of behavior reminiscent of polaron-host EuB$_6$ \cite{Das2012, brooks2004magnetic}. 
At low $T$ conventional 3D Heisenberg magnetism is observed. As the temperature increases,
a dramatic change in the  $\mu$SR relaxation rates occurs above $T \simeq 75$\,K, a temperature that coincides with the breakdown of the phenomenological DDH model in noise spectroscopy measurements. The latter is suggested to describe fluctuations of frozen clusters of linked magnetic polarons or magnetic domains in agreement with the observation of two magnetically distinct muon stopping sites coupled to the same magnetic order parameter.
Upon approaching $T_C$ and a maximum in the $1/f$-type noise a strong (power-law) increase of the resistance fluctuations and scaling behavior are observed, consistent with a percolation transition of magnetic polarons, similar to the CMR systems EuB$_6$ \cite{Das2012} and perovskite manganites \cite{Podzorov2000}. The unusal behavior of a two-level fluctuation process superimposed on the generic $1/f^\alpha$ noise in HgCr$_2$Se$_4$ can be described by a slowing down of the polaron dynamics, taking into account an effective radius of the magnetic polaron depending on the spin correlation length.
Above $T_C$, the $\mu$SR response suggests
that isolated magnetic polarons 
do not occupy a significant volume fraction of the sample.


\section*{Acknowledgements}
Work at Goethe-University Frankfurt was funded by the Deutsche Forschungsgemeinschaft (DFG, German Research Foundation) -- project number 449866704.
Muon measurements were made at the Swiss Muon Source and the STFC-ISIS Facility and we are grateful for the provision of beamtime. We are grateful to EPSRC (UK) (grant number: EP/N032128/1). M.~N.~Wilson acknowledges the support of the Natural Sciences and Engineering Research Council of Canada (NSERC).
Data from the UK effort will be made available via DOI:XXXXXXXX. M.~M.\ and J.~M.\ are grateful to Bernd Wolf for magnetic characterization measurements.

M.~M.\ and T.~J.~H.\ contributed equally to this work.  


\appendix
\section{Magnetic characterization}
\label{Appendix_magnetic}





\begin{figure}[h!]
\centering
\includegraphics[width=0.9\columnwidth]{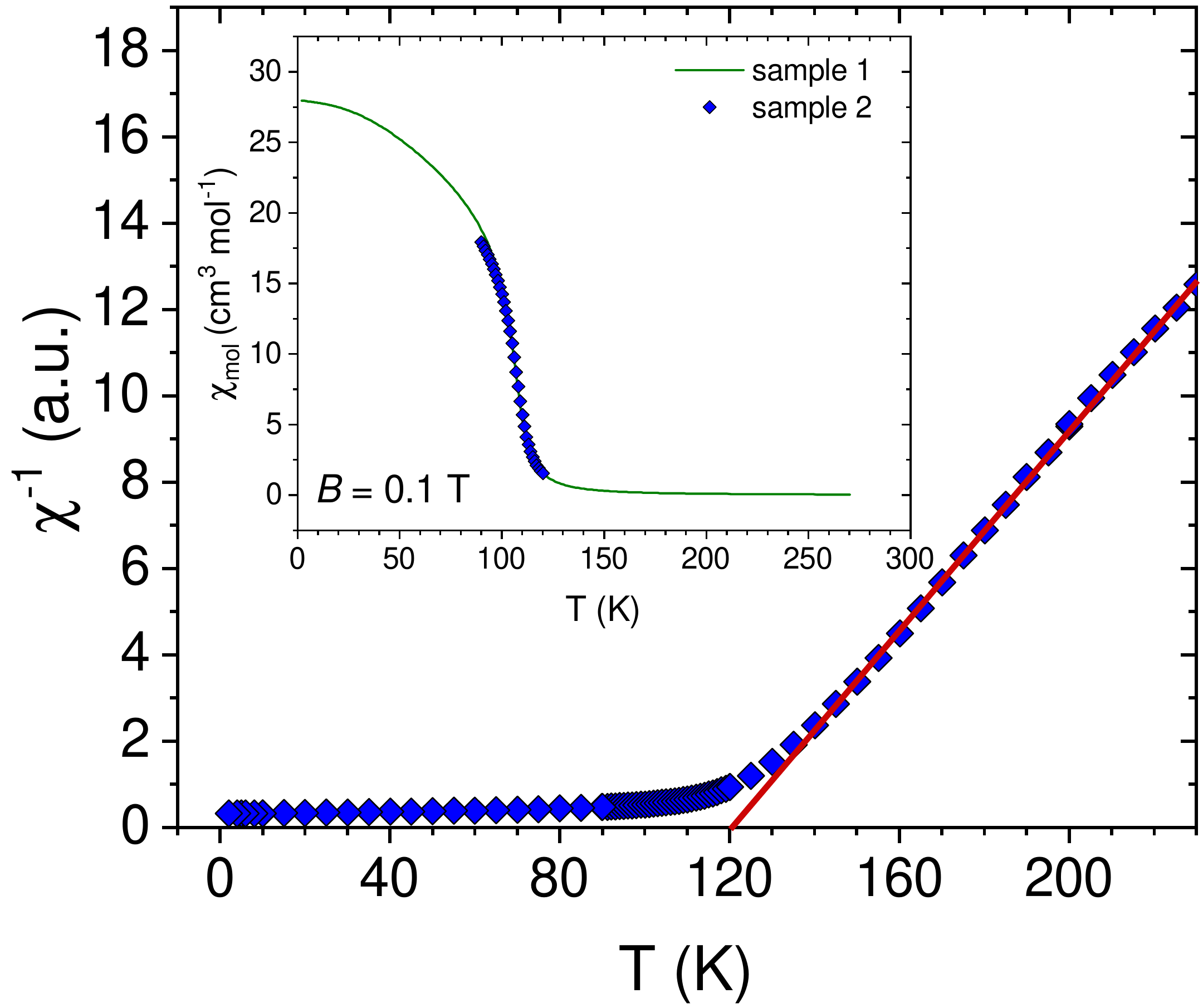}
\caption{Temperature dependence of the inverse magnetic susceptibility $1/\chi$ (blue diamonds) of $n$-HgCr$_2$Se$_4$ and the linear, Curie-Weiss  fit in the high temperature range (red line). The inset shows the temperature dependence of the magnetic susceptibility with an applied magnetic field $B$ = 0.1 \,T for different samples of the ferromagnetic semiconductor.} 
\label{sus}%
\end{figure}%
In Fig.\ \ref{sus} we plot the temperature dependence of the inverse magnetic susceptibility $1/\chi$ of $n$-HgCr$_2$Se$_4$. For the investigated sample, the susceptibility starts to deviate from the Curie-Weiss law (red line) at $T^* \approx$ 140 \,K. The inset shows the temperature dependence of the magnetic susceptibility at an applied magnetic field of $B = 0.1$\,T, showing a broadened magnetic transition.

\section{wTF $\mu$SR data}\label{sec:wtfmusr}

\begin{figure}[b]
	\centering
	\includegraphics[width=0.7\linewidth]{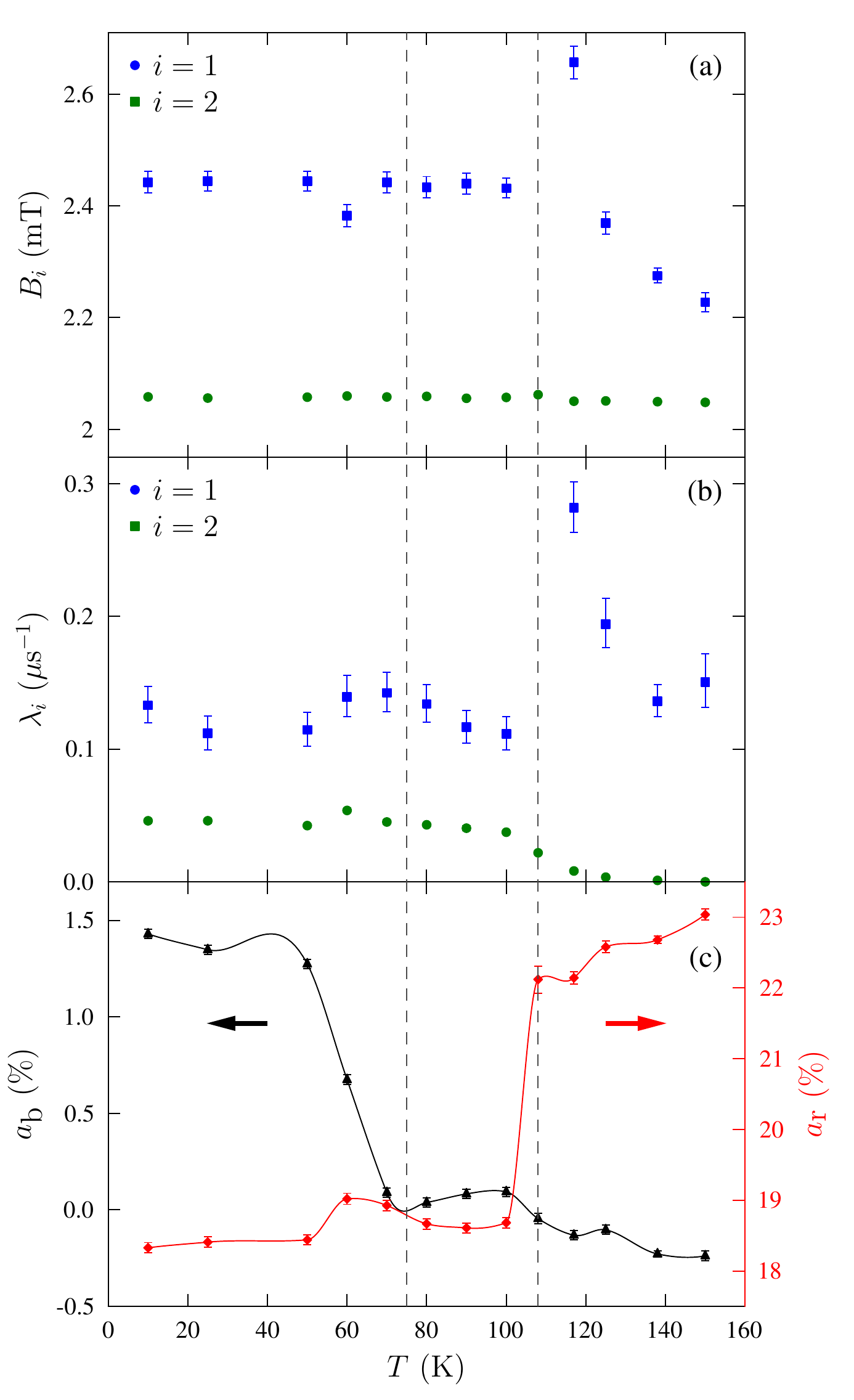}
	\caption{Parameters extracted through fitting wTF ($B_{\rm ext}~=~2$~mT) $\mu$SR measurements of HgCr$_2$Se$_4$ as described in the text. Dashed lines correspond to transitions between the three different regimes  identified in the text. Lines in (c) are a guide to the eye.}
	\label{fig:tf20}
\end{figure}
To investigate the possibility of magnetic phase separation, measurements were made at ISIS with a weak magnetic field ($B_{\rm ext}~=~2$~mT) applied perpendicular to the initial muon-spin polarisation (know as weak transverse field [wTF] measurement).
These measurements also allow us to investigate the changes in $a_{\rm b,r}$ above $T_{\rm c}$ shown in Fig.~\ref{fig:zf}.
At $T~<~T_{\rm c}$ the field $B_{\rm ext}$ is small compared to the internal fields, so only muons stopping outside of the sample will precess in this wTF.
There are therefore three anticipated contributions to $A\left(t\right)$: (i) muons that precess in the large internal field (which dephase too rapidly for detection at ISIS and therefore contribute a missing asymmetry, as already discussed), (ii) muons that have their initial spin along the internal field direction (expected to be 1/3 of the muon stopping in the sample in a polycrystalline sample) and therefore contribute a non-oscillating component $a_{\rm b}$, and (iii) muons that precess in $B_{\rm ext}$.
We therefore parametrize the wTF measurements using
\begin{equation}
	A\left(t\right) = a_{\rm r}\sum_{i=1}^{2}p_i\cos\left(\gamma_\mu B_i t + \phi_i\right)\exp\left(-\lambda_i t\right) + a_{\rm b} ,
\end{equation}
where $B_{1,2}$ are both found to be very close to $B_{\rm ext}~=~2$~mT.
The existence of two precessing components is likely due to the stray field of the sample causing a range of magnetic fields to be experienced by muons stopping outside of the sample.
The parameters extracted from these fits are shown in Fig.~\ref{fig:tf20}.

 As $T$ is reduced below $T_{\rm c}$, $a_{\rm r}$ is reduced as muons experience dynamic fluctuations on the muon-timescale, as already discussed, however $a_{\rm b}$ does not increase until below $T~\simeq~75$~K, where the 1/3-tail is recovered, once again identifying the three regions of behavior.
 These data suggest the freezing of dynamics might continue to occur down to lower temperatures  than implied by the change in dynamics seen in Fig.~\ref{fig:zf}.


Below $T_{\rm c}$ the total change of $a_{\rm b}$ is approximately 1.7\%, which should be 1/3 of the total change in $a_{\rm r}$ in a polycrystalline sample such as this, which is $5\%~\simeq~3\times1.7\%$.
This demonstrates that there is no missing asymmetry, and therefore no evidence for phase separation.
Above $T_{\rm c}$ there are only small changes, around 1\%, in $a_{\rm b,r}$; the increasing $a_{\rm r}$ suggests fewer muons are captured by the weak applied field.
This, along with the gradual collapse of the $B_1$ towards the applied field, suggests that we detect an effect due to fluctuating magnetic moments that dies away gradually above $T_{\rm c}$.

\section{Additional fluctuation spectroscopy data}
\label{Appendix_noise}
\begin{figure}[]
\centering
\includegraphics[width=0.9\columnwidth]{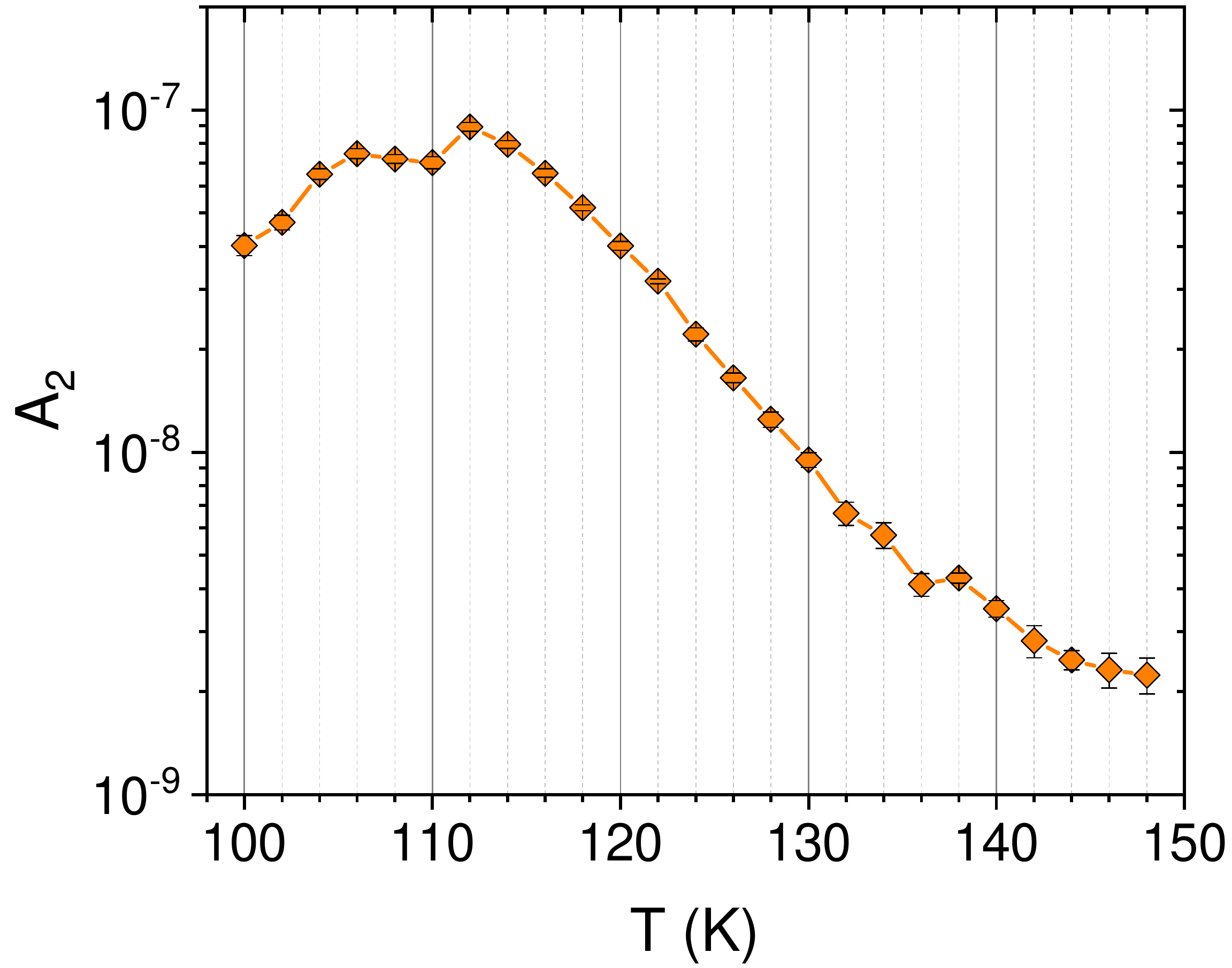}
\caption{Normalized resistance noise PSD, $S_R/R^2$ ($f$ = 1\,Hz) of the Lorentzian amplitude due to two-level switching processes observed in the grey shaded region between 100 and 150 \,K. }
\label{Lorentzian}%
\end{figure}%
In Fig.\ \ref{Lorentzian} we show the magnitude of the Lorentzian part of the noise spectra $A_2(T)$ in the grey shaded area between 100 and 150\,K. Starting at 150\,K the power spectral density starts to increase, consistent with the overall noise amplitude of the $1/f$ noise, $A_1(T)$, shown in Fig.\ \ref{noise}. At a temperature of 112\,K, coinciding with the deviations seen in the percolation modelling, the noise magnitude of A$_2(T)$ begins to flatten and decrease upon further lowering the temperature.  

\begin{figure}[b]
\centering
\includegraphics[width=0.9\columnwidth]{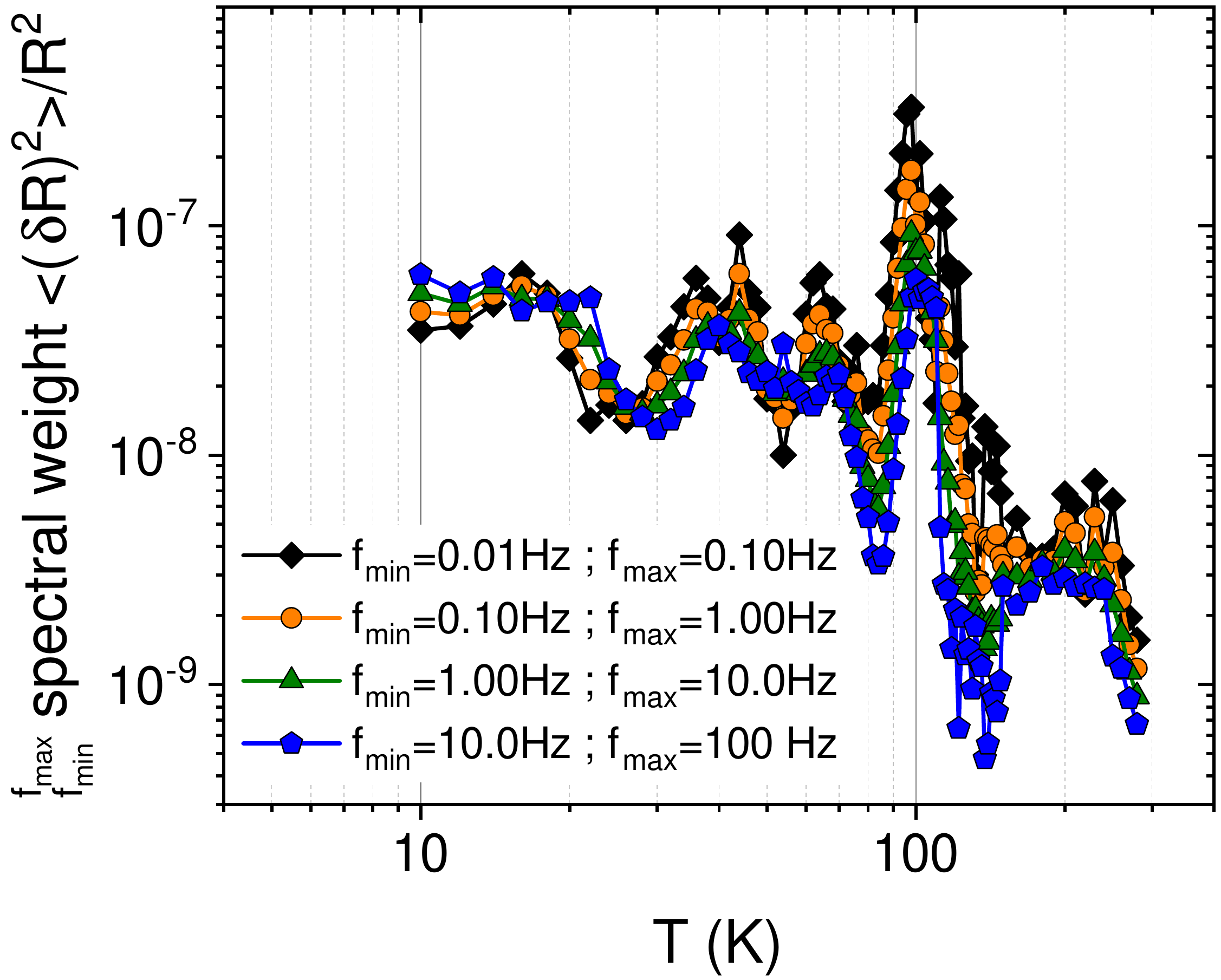}
\caption{Calculated spectral weight of the normalized resistance fluctuations $\langle (\delta R)^2 \rangle/R^2$ for four different frequency decades as a function of temperature $T$.} 
\label{weight}%
\end{figure}

Following the phenomenological DDH model, shown in fig. \ref{DDH} (a), a frequency exponent $\alpha > 1$ above 1 means a domination of slow fluctuations whereas for $\alpha < 1$ fluctuators with higher energies contribute more which means that slower processes dominate the dynamics. 
In Fig.\ \ref{weight} we present the calculated spectral weight in different frequency windows. 
Starting at room temperature slower dynamics has a larger amplitude, and upon by approaching the percolation transition at $\sim 98$\,K those processes become even more dominant in agreement with the behavior of the distinct two-level fluctuator and the importance of spin correlations. 



\begin{thebibliography}{99}


\bibitem{Baltzer1966} P. K. Baltzer, P. J. Wojtowicz, M. Robbins, and E. Lopatin, 
Phy. Rev. \textbf{151}, 367 (1966).

\bibitem{Guan2015}
T. Guan, C. Lin, C. Yang, Y. Shi, C. Ren, Y. Li, H. Weng, X. Dai, Z. Fang, S. Yan, and P. Xiong, 
Phys. Rev. Lett. \textbf{115}, 087002 (2015).

\bibitem{Yang2019}
S. Yang, Z. Li, C. Lin, C. Yi, Y. Shi, D. Culcer, and Y. Li, 
Phys. Rev. Lett. \textbf{123}, 096601 (2019).

\bibitem{Lin2016}
C. Lin, C. Yi, Y. Shi, L. Zhang, G. Zhang, J. M\"uller, and Y. Li, 
Phys. Rev. B \textbf{94}, 224404 (2016).

\bibitem{Molnar2007} S. von Moln\'{a}r and P. A. Stampe, {\it Magnetic Polarons}, in Handbook of Magnetism and Advanced
Magnetic Materials, eds. H. Kronm\"uller and S. Parkin, Vol. 5, John Wiley \& Sons, Ltd. (2007).

\bibitem{Majumdar1998} P. Majumdar and P. Littlewood, Phys. Rev. Lett. {\bf 81}, 1314, (1998).

\bibitem{Teresa1997} J. M. De Teresa et al., Nature {\bf 386}, 256 (1997).

\bibitem{Faeth1999} M. F\"ath et al., Science {\bf 285}, 1540 (1999).

\bibitem{Adams2000} C. P. Adams et al., Phys. Rev. Lett. {\bf 85}, 3954 (2000).

\bibitem{Tokura2006} Y. Tokura, Rep. Prog. Phys. {\bf 69}, 797 (2006).

\bibitem{Urbano2004} R. R. Urbano, P. G. Pagliuso, C. Rettori, S. B. Oseroff, J. L. Sarrao, P. Schlottmann, and Z. Fisk, Phys. Rev. B {\bf 70}, 140401(R) (1998).

\bibitem{Nyhus1997} P. Nyhus et al., Phys. Rev. B {\bf 56}, 2717 (1997).

\bibitem{Suellow1998} S. S\"ullow et al., Phys. Rev. B 57, 5860 (1998).

\bibitem{Suellow2000} S. S\"ullow et al, Phys. Rev. B 62, 11626 (2000).

\bibitem{Yu2006} U. Yu, B. I. Min, Phys. Rev. B {\bf 74}, 094413 (2006).

\bibitem{Zhang2009} X. H. Zhang et al., Phys. Rev. Lett. {\bf 103}, 106602 (2009).

\bibitem{Li2018}
L. Li, L.-Q. Yan, Y. Shi, P. Lu, and Y. Sun, 
Journal of Physics: Condensed Matter \textbf{30}, 144804 (2018).

\bibitem{Yang2004} Z. Yang, X. Bao, S. Tan, and Y. Zhang, Phys. Rev. B {\bf 69}, 144407 (2004).

\bibitem{Xia2019} Q. S. Xia, J. Li, C. N. Wang, and J. Wen, Low Temp. Phys. {\bf 45}, 1076 (2019).

\bibitem{Arai1973} 
T. Arai, M. Wakaki, S. Onari, K. Kudo, T. Satoh, and T. Tsushima, 
Journal of the Physical Society of Japan, 
\textbf{34}, 68 (1973).

\bibitem{Wang2013}
W. D. Wang, A. Li, T. Dong, M. Lei, X. L. Fu, S. S. Miao, P. Zheng, P. Wang, G. Y. Shi, J. L. Luo, and N. L. Wang, 
Journal of Low Temperature Physics \textbf{171}, 127-134 (2013).

\bibitem{blundell1999spin}
S. J. Blundell, Spin-polarized muons in condensed matter physics, Contemporary Physics \textbf{40}, 175 (1999).

\bibitem{blundell2021}
S. J. Blundell, R. De Renzi, T. Loncaster, and F. L. Pratt, OUP, Oxford (2021).

\bibitem{pratt2000wimda}
F. L. Pratt, WiMDA: a muon data analysis program for the Windows PC, Physica B: Condensed Matter \textbf{289}, 710 (2000).

\bibitem{james1975minuit}
F. James and M. Roos, Minuit: a system for function minimization and analysis of the parameter errors and corrections, Comput. Phys. Commun. \textbf{10}, 343 (1975).

\bibitem{iminuit}
H. Dembinski, P. Ongmongkolkul, C. Deil, D. Menendez Hurtado, H. Schreiner, M. Feickert, Andrew, C. Burr, F. Rost, A. Pearce, L. Geiger, B. M. Wiedemann, Gonzalo, M. Gorelli, and O. Zapata, iminuit – a Python interface to Minuit, accessed: 02-02-2021.

\bibitem{Mueller2011}
J. M\"uller, 
ChemPhysChem \textbf{12}, 1222 (2011).

\bibitem{Mueller2018}
J. M\"uller and T. Thomas, 
Crystals \textbf{8} (2018).

\bibitem{pelissetto2002critical}
A. Pelissetto and E. Vicari, Critical phenomena and renormalization-group theory, Physics Reports \textbf{368}, 549 (2002).

\bibitem{pratt2007chiral}
F. L. Pratt, P. J. Baker, S. J. Blundell, T. Lancaster, M. A. Green, and M. Kurmoo, Chiral-like critical behavior in the antiferromagnet cobalt glycerolate, Physical Review Letters \textbf{99}, 017202 (2007).

\bibitem{pospelov2019non}
E. A. Pospelov, V. V. Prudnikov, P. V. Prudnikov, and A. S. Lyakh, Non-equilibrium critical behavior of the 3D classical Heisenberg model, Journal of Physics: Conference Series \textbf{1163}, 012020 (2019).

\bibitem{Solin2008}
N. I. Solin, V. V. Ustinov, and S. V. Naumov, 
Physics of the Solid State \textbf{50}, 901 (2008).


\bibitem{Podzorov2000}
V. Podzorov, M. Uehara, M. E. Gershenson, T. Y. Koo, and S.-W. Cheong, 
Phys. Rev. B \textbf{61}, R3784 (2000).

\bibitem{Das2012}
P. Das, A. Amyan, J. Brandenburg, J. M\"uller, P. Xiong, S. Molnar and Z. Fisk, 
Phys. Rev. B. \textbf{86}, 184425 (2012).

\bibitem{Bogdanovic2002} S. Bogdanovich and D. Popovi${\rm \acute{c}}$, Phys. Rev. Lett. \textbf{88}, 236401 (2002).

\bibitem{Kar2003} S. Kar, A. K. Raychaudhuri, A. Ghosh, H. von L\"ohneysen, and G. Weiss, Phys. Rev. Lett. \textbf{91}, 216603 (2003).

\bibitem{Hartmann2015} B. Hartmann,1 D. Zielke, J. Polzin, T. Sasaki, and J. M\"uller, Phys. Rev. Lett. \textbf{114}, 216403 (2015).

\bibitem{Kogan1996}
Sh. Kogan,
\textit{Electronic Noise and Fluctuations in Solids}, 
Cambridge University Press, Cambridge, 1996.

\bibitem{Tremblay1985}
A.-N. S. Tremblay, S. Feng, and P. Breton, 
Phys. Rev. B. \textbf{33}, 2077-2080 (1985).

\bibitem{Rudman1985}
D. A. Rudman, J. J. Calabres, and J. C. Garland, 
Phys. Rev. B \textbf{33}, 1456-1459 (1985).

\bibitem{Bhatt2002}
R. N. Bhatt, M. Berciu, M. P. Kennett, and X. Wan, 
Journal of Superconductivity
\textbf{1}, 71 (2002).

\bibitem{Kaminski2002}
A. Kaminski, and S. Das Sarma, 
Phys. Rev. Lett. \textbf{88}, 247202 (2002).

\bibitem{comment3}
Note that the lowest data points are not included in the fit. This may be justified by sample inhomogeneities which also cause the dip in $\rho(T)$ and the split peak in the $1/f$-noise magnitude at $\sim 100$\,K. For the fit of the data in Fig.\ \ref{corner-frequency} to the model Eq.\ (\ref{model}) the following parameters are used. The exponent $\nu = 0.709$ is given in the 3D Heisenberg model and the localization (Bohr) radius of the donor $a_B = 0.53 \epsilon/(m^\ast/m_e)\,{\rm \AA}$
can be fixed to $a_B = 4.6$\,nm by using $\epsilon = 13\,\epsilon_0$ and $m^\ast = 0.15\,m_e$ \cite{Yang2019}. The parameter $\ln{f_0} = 3.69$ can be read of the figure by estimating the nearly constant values at higher temperatures. The fitting yields $\xi_0 = 0.6$\,nm, $J_0 = 173$\,K (16\,meV), $K = 20$\,K/nm$^3$ and $T_C = 98$\,K. For $T_C$ it was necessary not to choose the bulk ferromagnetic transition temperature but rather a lower temperature close to the maximum of the $1/f$-noise (and the resistivity) as a starting value for the fit procedure. This also is in agreement with a rather broad magnetic transition and sample inhomogeneities. We point out that due to the large number of free parameters the fit is not unique. 

\bibitem{Raquet2001}
B. Raquet, 
\textit{Electronic Noise in Magnetic Materials and Devices}, 
Springer Berlin Heidelberg, Berlin Heidelberg, 2001.

\bibitem{Pre1950}
F. K. Du Pr${\rm \acute{e}}$, 
Phys. Rev. B \textbf{78}, 615 (1950).

\bibitem{comment2} 
The correlation function of a two-level process decays exponentially with a characteristic time $\tau_c$. The superposition of many such processes, however, results in a non-exponential dynamics.

\bibitem{Dutta1979}
P. Dutta, P. Dimon, and P. M. Horn,
Phys. Rev. Lett. \textbf{43}, 646-649 (1979).

\bibitem{Raquet1999}
B. Raquet, J. M. D. Coey, S. Wirth, and S. Moln${\rm \acute{a}}$r,
Phys. Rev. B \textbf{59}, 12435-12443 (1999).

\bibitem{brooks2004magnetic}
M. L. Brooks, T. Lancaster, S. J. Blundell, W. Hayes, F. L. Pratt, and Z. Fisk, Magnetic phase separation in EuB$_6$ detected by muon spin rotation, Physical Review B \textbf{70}, 020401 (2004).


\bibitem{fisk1979magnetic} Z. Fisk, D. Johnston, B. Cornut, S. Von Molnar, S. Os- eroff, and R. Calvo, Magnetic, transport, and thermal properties of ferromagnetic EuB6, Journal of Applied Physics {\bf 50}, 1911 (1979).

\bibitem{Pohlit2018}
M. Pohlit, S. R\"o{\ss}ler, Y. Ohno, H. Ohno, S. von Moln\'{a}r, Z. Fisk, J. M\"uller, and S. Wirth,
Phys. Rev. Lett. \textbf{120}, 257201 (2018).

\end{thebibliography}
\end{document}